\documentclass[prapplied,superscriptaddress,notitlepage]{revtex4-1}
\usepackage{soul}
\usepackage{amsthm}
\usepackage{amsmath}
\usepackage{amssymb}
\usepackage{graphicx}
\usepackage{rotating}
\usepackage{epstopdf}
\usepackage{subfig}
\usepackage{multirow}
\usepackage{array}
\usepackage{adjustbox}
\usepackage{makecell}
\usepackage{url}
\usepackage{float}
\usepackage{bm}
\usepackage{ifthen}
\usepackage[usenames,dvipsnames]{color}
\usepackage{colortbl}
\usepackage{mathrsfs}
\usepackage[colorlinks=false,citecolor=blue,urlcolor=black]{hyperref}

\newcommand{\tn}[1]{\textnormal{#1}}
\newcommand{\be}{\begin{equation}}
\newcommand{\ee}{\end{equation}}

\newcommand{\rv}[1]{{\bf{#1}}}

\usepackage{placeins}
\newcommand{\sket}[1]{{\ensuremath{\lvert#1\rangle}}}
\newcommand{\lket}[1]{{\ensuremath{\left\lvert#1\right\rangle}}}
\newcommand{\ket}[1]{\if@display\lket{#1}\else\sket{#1}\fi}

\newcommand{\sbra}[1]{{\ensuremath{\langle#1\rvert}}}
\newcommand{\lbra}[1]{{\ensuremath{\left\langle#1\right\rvert}}}
\newcommand{\bra}[1]{\if@display\lbra{#1}\else\sbra{#1}\fi}

\newcommand{\sbraket}[2]{{\ensuremath{\langle#1\rvert#2\rangle}}}
\newcommand{\lbraket}[2]{{\ensuremath{\left\langle#1\!\left\rvert\vphantom{#1}#2\right.\!\right\rangle}}}
\newcommand{\braket}[2]{\if@display\lbraket{#1}{#2}\else\sbraket{#1}{#2}\fi}
\newcommand{\bracket}[2]{\braket{#1}{#2}}

\newcommand{\sketbra}[2]{{\ensuremath{\lvert #1\rangle\!\langle #2\rvert}}}
\newcommand{\lketbra}[2]{{\ensuremath{\left\lvert #1\right\rangle\!\!\left\langle #2\right\rvert}}}
\newcommand{\ketbra}[2]{\if@display\lketbra{#1}{#2}\else\sketbra{#1}{#2}\fi}

\newcommand{\eps}{\varepsilon}

\theoremstyle{plain}

\theoremstyle{definition}

\begin{document}
\title{Hyperentangled Time-bin and Polarization Quantum Key Distribution}
\author{Joseph C. Chapman}
\email{chapmanjc@ornl.gov}
\affiliation{Oak Ridge National Laboratory, Oak Ridge, TN 37831}
\affiliation{Illinois Quantum Information Science and Technology Center, University of Illinois at Urbana-Champaign, Urbana, IL 61801}
\affiliation{Department of Physics, University of Illinois at Urbana-Champaign, Urbana, IL 61801}
\author{Charles C. W. Lim}
\affiliation{Department of Electrical \& Computer Engineering, National University of Singapore, Singapore 117583}
\affiliation{Centre for Quantum Technologies, National University of Singapore, Singapore 117583}
\author{Paul G. Kwiat}
\affiliation{Illinois Quantum Information Science and Technology Center, University of Illinois at Urbana-Champaign, Urbana, IL 61801}
\affiliation{Department of Physics, University of Illinois at Urbana-Champaign, Urbana, IL 61801}

%\date{}
\begin{abstract}

{Fiber-based quantum communication networks are currently limited without quantum repeaters. Satellite-based quantum links have been proposed to extend the network domain. We have developed a quantum communication system, suitable for realistic satellite-to-ground communication. With this system, we have executed an entanglement-based quantum key distribution (QKD) protocol developed by Bennett, Brassard, and Mermin in 1992 (BBM92), achieving quantum bit error rates (QBER) below 2$\%$ in all bases. More importantly, we demonstrate low QBER execution of a higher dimensional hyperentanglement-based QKD protocol, using photons simultaneously entangled in polarization and time-bin, leading to significantly higher secure key rates, at the cost of increased technical complexity and system size. We show that our protocol is suitable for a space-to-ground link, after incorporating Doppler shift compensation, and verify its security using a rigorous finite-key analysis. Additionally, We discuss system engineering considerations relevant to those and other quantum communication protocols, and their dependence on what photonic degrees of freedom are utilized.}
\end{abstract}
\maketitle
%\tableofcontents
\section{Motivation and Background}
Implementing quantum communication protocols such as quantum key distribution (QKD) over long distances is a major step toward addressing the challenge of establishing a global quantum network. To lay dedicated dark fiber over long distances is expensive and non-reconfigurable, and, without quantum repeaters, such links have very low transmission, due to the exponential decrease in fiber transmission with distance. It has been proposed to instead use space-based links with free-space quantum channels between a ground station and an orbiting platform~\cite{spaceqcom,aspelmeyer2003long}, or to combine these with fiber links~\cite{simon2017towards}. Such a channel has much lower loss than fiber over the same distance---transmission drops only quadratically, due to diffraction---allowing much more efficient protocol execution over comparable distances. For example, the recent achievement of entanglement distribution from a satellite to two ground stations realized a loss reduction of some 12 orders of magnitude~\cite{entanglement2017sat}; the same satellite also demonstrated decoy-state QKD, realizing a 20 orders-of-magnitude enhancement in channel transmission~\cite{qkd2017sat,liao2018satellite}. This satellite was used to implement the 1992 entanglement-based QKD protocol by Bennett, Brassard, and Mermin (BBM92)~\cite{BBM92,yin2017satellite}, though the detection rate and signal-to-noise ratio were far too low to generate secret key of any substantial length (finite-key effects were not rigorously accounted for in the original analysis, but now have been in an updated analysis with slightly higher security error~\cite{lim2021security}). A number of other groups around the world are also working on similar endeavors~\cite{grieve2018spooqysats,vallone2015experimental,pugh2017airborne,steinlechner2017distribution}. 

To this same end, we have developed a single system to implement multiple quantum protocols relevant for satellite-based communication. We have previously characterized the performance of this system to implement superdense teleportation~\cite{chapman2019time,chapphowest2018},  and have used it to demonstrate high-dimensional tests of nonlocality~\cite{zeitlerbell}. Given that our system already generates entanglement and hyperentanglement (simultaneous entanglement in multiple degrees of freedom~\cite{PhysRevLett.95.260501}), here we consider only entanglement-based protocols, which in any event offer advantages in terms of protection from side-channel attacks. In particular, we implement polarization-entanglement-based QKD (specifically, the BBM92 protocol~\cite{BBM92}) as well as a novel higher dimensional, hyperentanglement-based QKD protocol (HEQKD). We include a finite-key security analysis and secret-key-rate simulation of both protocols (see Appendix \ref{FKA} for full analysis) and characterize our lab-based implementation of both protocols under conditions relevant for a satellite-to-Earth implementation. Using our finite-key analysis, we simulate a currently feasible upgraded version of our system, to directly compare the projected performance of BBM92 and HEQKD in a space-to-Earth channel. Note that while several recent articles have analyzed space-to-Earth or ground-to-space quantum key distribution \cite{PhysRevApplied.16.014067,pugh2020adaptive,villasenor2020atmospheric,sharma2019analysis}, none have included a comparative study incorporating hyperentanglement. Finally, we include a discussion of the impact of quantum communication between rapidly moving platforms using various degrees of freedom, e.g., polarization and time bin, identifying requirements for each.

\section{Protocol Introduction}
\label{sec:ProtIntro}
\subsection{Qubit-Entangled BBM92}
For BBM92~\cite{BBM92}, Alice's and Bob's bases are written as $\{\mathsf{A}_i\}_{i=1}^2$ and $\{\mathsf{B}_i\}_{i=1}^2$, respectively. Assuming Alice and Bob are each operating in a two-dimensional Hilbert space with computational basis given by $\mathsf{Z}=\{\ket{j}\}_{j=0}^1$; their measurement bases are defined as $\mathsf{A}_1\equiv\mathsf{Z}$, $\mathsf{A}_2\equiv\{(\ket{0}\pm\ket{1})/\sqrt{2} \}$, and similarly for Bob, where here $\ket{0}\equiv\ket{H}$, $\ket{1}\equiv\ket{V}$ for $H$, $V$ representing horizontal and vertical polarization, respectively. We assume the bases are uniformly chosen with probability 1/2 and the key is generated from both bases. Here, the size of the raw key is denoted by $m=n+k$, where $k=m(1-r)$ is the amount of raw key used for parameter estimation, $n=mr$ is the amount of raw key left for key generation, and $r$ is the parameter estimation ratio. 

The BBM92 protocol ideally requires Alice and Bob to receive photons from a maximally entangled photon-pair source. In our case, 1550-nm and 810-nm photons are entangled in their polarization (see Sect. \ref{expsetupsec}), and the state shared between Alice and Bob is
\begin{equation}
\ket{\Psi_{AB}}=\frac{1}{\sqrt{2}}\ket{t_1^pt_1^p}\otimes (\ket{H_{1550}H_{810}}+\ket{V_{1550}V_{810}})\text{,}
\end{equation}
where $t_1^p$, represents pump time bin ``1".
\subsection{HEQKD}
For HEQKD, Alice's and Bob's bases are written as $\{\mathsf{A}_i\}_{i=1}^4$ and $\{\mathsf{B}_i\}_{i=1}^4$, respectively. We assume that Alice and Bob are each operating in a four-dimensional Hilbert space with the computational basis given by $\mathsf{Z}=\{\ket{j}\}_{j=0}^3$; their measurement bases are defined as $\mathsf{A}_1\equiv\mathsf{Z}$, $\mathsf{A}_2\equiv\{(\ket{0}\pm\ket{1})/\sqrt{2},(\ket{2}\pm\ket{3})/\sqrt{2} \}$, $\mathsf{A}_3\equiv\{(\ket{0}\pm\ket{2})/\sqrt{2},(\ket{1}\pm\ket{3})/\sqrt{2} \}$, and $\mathsf{A}_4\equiv\{(\ket{0}+\ket{1}+\ket{2}-\ket{3})/2$, $(\ket{0}+\ket{1}-\ket{2}+\ket{3})/2$, $ (\ket{0}-\ket{1}+\ket{2}+\ket{3})/2$, \\$ (\ket{0}-\ket{1}-\ket{2}-\ket{3})/2 \}$, and similarly for Bob. Not that, while others have studied higher dimensional quantum key distribution~\cite{islam2017provably,Eckerhyperentang,mirhosseini2015high, bouchard2018experimental}, to the best of our knowledge no previous work has both used hyperentanglement and been able to make the full set of measurements required to implement higher dimensional QKD, i.e., measurements in a pair of fully mutually-unbiased bases. In our implementation, we realize the higher-dimensional encoding by using polarization ($\ket{H}$ and $\ket{V}$) and one time-bin qubit (with basis states $\ket{t_1^p}$ and $\ket{t_2^p}$, the two time-bins for the pump): $\ket{0}\equiv\ket{Ht_1^p}$, $\ket{1}\equiv\ket{Vt_2^p}$, $\ket{2}\equiv\ket{Vt_1^p}$, and $\ket{3}\equiv\ket{Ht_2^p}$. It can be easily checked that bases 1 and 4 and bases 2 and 3 are mutually unbiased. Bases 1 and 2 are each chosen with probability $p$, while bases 3 and 4 are each chosen with probability $q$. Hence, we have that $2p+2q=1$, or $q=1/2-p$.

HEQKD provides more raw bits of key per photon (2 instead of 1) and higher error resilience than BBM92~\cite{NDQKD}, but requires generating and distributing hyperentangled photons~\cite{PhysRevLett.95.260501}. As shown in Sect. \ref{expsetupsec}, we prepare non-degenerate hyperentangled photons (at 1550 nm and 810 nm) that are entangled in their polarization and time-bin degrees of freedom:
\begin{align}
\ket{\Psi_{AB}}= \frac{1}{2}[&\ket{(Ht_1^p)_{810}(Ht_1^p)_{1550}}+\ket{(Vt_2^p)_{810}(Vt_2^p)_{1550}}+\notag\\
&\ket{(Vt_1^p)_{810}(Vt_1^p)_{1550}}+\ket{(Ht_2^p)_{810}(Ht_2^p)_{1550}}]\text{.} \label{psiAB}
\end{align}
\section{Security Analysis and Secret Key Simulation Summary}
In the following, we summarize the finite-key security of BBM92 (also called entanglement-based BB84) and our high-dimensional QKD protocol using two sets of two mutually unbiased bases. The security proof technique relies on the entropic uncertainty relation~\cite{tightanalysis}; see Appendix \ref{FKA} for the full analysis. In our analysis, we trust the local measurements, but the entanglement source can be untrusted.

In both above protocols, we assume that entanglement is generated in Alice's laboratory using a non-deterministic photon pair source, e.g., a spontaneous parametric down-conversion (SPDC) or four-wave mixing source. Using standard models for that type of source~\cite{Ma2007}, if the probability of $n$ photon-pairs being generated is $P_n$ and the coincidence detection probability of $n$ photon-pairs is $Y_n$, we calculate the overall coincidence detection probability per pump pulse using non-photon-number-resolving detectors to be
\be
R=\sum_{n=0}^\infty P_nY_n = 1- \frac{1-\xi_A}{(1+\eta_A\gamma)^2}-\frac{1-\xi_B}{(1+\eta_B\gamma)^2}+\frac{(1-\xi_A)(1-\xi_B)}{(1+\eta_A\gamma + \eta_B\gamma-\eta_A\eta_B\gamma)^2},
\ee
where $\gamma$ is related to the pump power of the laser ($2\gamma=\mu$ is the mean number of pairs per pump pulse), $\xi_A$ ($\xi_B$) is the probability of observing a noise or background count on Alice's (Bob's) side per pump pulse, and $\eta_A$ ($\eta_B$) is the overall detection efficiency for Alice (Bob). For encoding in a $d$-dimensional quantum state, our estimated error probability per pump pulse is
\be
Q_{d,\tn{est}}=\frac{E_{b}+E_{\ket{\Phi^n}}+E_{\text{MPE}}}{R},\label{qberest}
\ee
where $d=2$ for BBM92, $d=4$ for HEQKD, $E_{b}$ is the estimated probability of noise or a background count causing an error, $E_{\ket{\Phi^n}}$ is the probability of errors when all photons produced by SPDC are detected, and $E_{\text{MPE}}$ is the estimated probability of errors when at least one pair of photons was detected from an $n$-photon-pair state, not including terms when all photons produced by SPDC are detected (See Appendix \ref{FKA} for details).

In our analysis of both protocols we use the standard security definitions for QKD~\cite{composability}: we say that the QKD protocol is $\eps$-secure if it is both $\eps_\tn{sec}$-secret and $\eps_\tn{cor}$-correct, where $\eps=\eps_\tn{sec}+\eps_\tn{cor}$ (see Appendix \ref{FKA} for full definition). Using the quantum leftover-hash lemma~\cite{tomhayinfo} and the entropic uncertainty relation~\cite{tightanalysis}, we can bound the min-entropy and, therefore, the secret key length $\ell^{2D}$ of BBM92:
\be
\ell^{2D} = \max_{\beta \in (0,\eps_\tn{sec}/4)} \left\lfloor n_\tn{ext}^{2D} +4\log_2 \beta -2 \right\rfloor, 
\ee
where
\begin{equation}
n_\tn{ext}^{2D}\equiv n(1-h_2(Q_{2,\tn{est}}+\Delta(n,k,\beta))) -\tn{Leak}_{\tn{EC}}^{2D}-\log_2\frac{2}{\eps_\tn{cor}}
\end{equation}
is the extractable key length, $h_2$ is the binary entropy (Eqn. \ref{binaryentropy}), $\Delta$ is the statistical noise due to finite statistics of the amount of raw key used for parameter estimation $k$ and the raw key left for key generation $n$, $\tn{Leak}_{\tn{EC}}^{2D}$ is any information leaked during error correction, and $\beta$ is a constrained parameter to be optimized~\cite{tomhayinfo}.

In our analysis of HEQKD, we focus on processing Alice's measurement data to extract her secret key; our analysis can easily be converted for Bob's measurement data. Specifically, we extract key when Alice measures in basis 1 (a random string of length $n_1=m_{1,1}+m_{1,2}+m_{1,3}$, where, e.g., $m_{1,2}$ are the measurements in basis 1 for Alice and basis 2 for Bob) and basis 2 (a random string of length $n_2=m_{2,1}+m_{2,2}+m_{2,4}$). The measurements in basis 3, $m_{3,3}$, and basis 4, $m_{4,4}$, are used for parameter estimation, i.e., eavesdropper detection, of basis 2 and 1, respectively. The other basis combinations are unusable. Again, using the quantum leftover-hash lemma~\cite{tomhayinfo} and a version of the entropic uncertainty relation for $d=4$ mutually unbiased bases, after some simplification we find the secret key length:
\be
\ell^{4D} = \max_{\beta \in (0,\eps_\tn{sec}/4)} \left\lfloor n_\tn{ext}^{4D}+4\log_2 \beta -2 \right\rfloor, 
\ee
where
\begin{multline}
n_\tn{ext}^{4D}\equiv n_1(2 - h_4(Q_{4,4,4,\tn{est}}+\nu(n_1,m_{4,4},\bar{\eps})))+n_2(2 - h_4(Q_{3,3,4,\tn{est}}+\nu(n_2,m_{3,3},\bar{\eps})))-\tn{Leak}_{\tn{EC}}^{4D}-\log_2\frac{2}{\eps_\tn{cor}}
\end{multline} 
is the \emph{extractable} key length, $h_4$ is the Shannon entropy (Eqn. \ref{shannonentropy}), $Q_{i,i',4,\tn{est}}$ is the observed error rate conditioned on Alice and Bob respectively choosing basis $i$ and $i'$, $\bar{\eps}\equiv\eps_\tn{sec}/6-\beta/3$, $\nu$ is the statistical noise due to finite statistics, and $\tn{Leak}_{\tn{EC}}^{4D}$ is any information leaked during error correction (See Appendix \ref{FKA} for details).

\section{Experimental Setup}\label{expsetupsec}
\begin{figure}
\centerline{\includegraphics[scale=0.65]{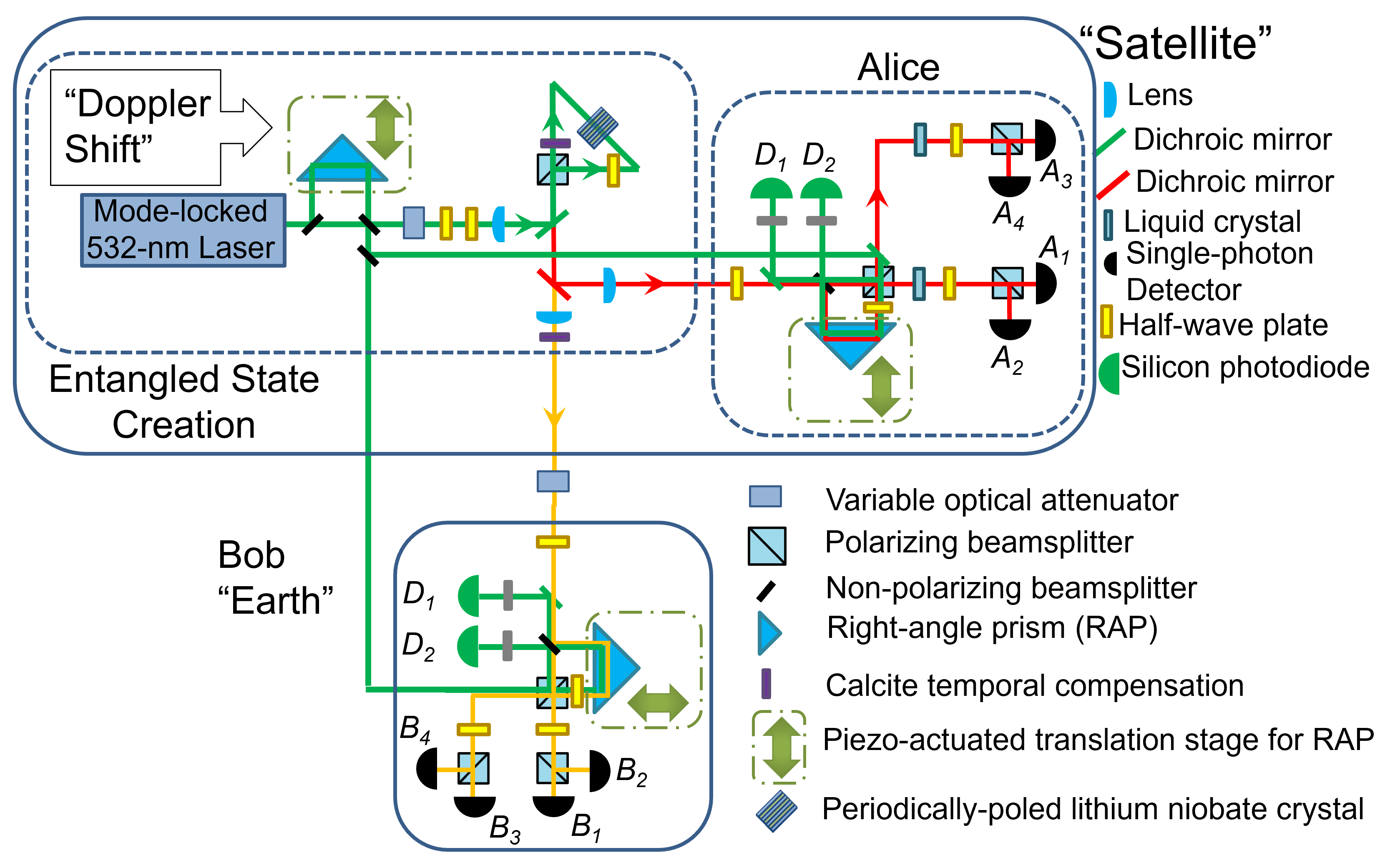}}\caption{BBM92 and HEQKD Optical Setup: Photonic ``ququarts'' hyperentangled in polarization and time-bin are generated via spontaneous parametric down-conversion in periodically poled lithium niobate. Green lines are the 532-nm pump (and stabilization) beam; red and yellow are the signal (810 nm) and idler (1550 nm) photons, respectively. For BBM92, the pump right-angle prism was blocked so there was no time-bin entanglement. In both Alice's or Bob's analyzer, the short arm measured photons in the H/V basis and the long arm measured photons in the D/A basis (using a half-wave plate, just before the polarizing beamsplitter, set at 22.5$^\circ$ from Horizontal). For HEQKD, both time-bin and polarization entanglement are used; see Tables \ref{BBM92detmapping} and \ref{HEQKDdetmapping} for the measurement-to-detector mapping. The phase in the phase-sensitive bases (bases 2, 3, and 4) is tuned by tilting the half-wave plate before the Sagnac interferometer and/or actuating the liquid crystals after Alice's analyzer interferometer. All half-wave plates just before the detectors were set at 22.5$^\circ$ from horizontal. The half-wave plate before the analyzer interferometer was set at 0 (22.5$^\circ$) from horizontal for measurements in bases 1,2 (3,4).}\label{heqkdsetup}
\end{figure}
To generate entangled photons in time-bin and polarization, we use an 80-MHz mode-locked, 532-nm laser (Spectra Physics Vanguard 2.5W 355 laser), frequency doubled from 1064 nm, with a pulse width of about 7 ps. Each pulse of this beam is split into two time-bins using about a 2.4-ns delay (see Fig. \ref{heqkdsetup}). The beam is then used to pump a type-0 periodically poled lithium niobate (PPLN) crystal (poling period of 7.5 $\mu$m) inside a polarizing Sagnac interferometer. This Sagnac entangled photon source~\cite{Sagnac1,Sagnac2}, ignoring time bins, produces the state 
\begin{equation}
\frac{\big(\ket{H}_{810}\ket{H}_{1550}+e^{i\phi}\ket{V}_{810}\ket{V}_{1550}\big)}{\sqrt{2}}\text{.}
\end{equation}
The 532-nm pump has a bandwidth of 64 GHz full-width at half-maximum (FWHM). The peaks (FWHM bandwidths) of the down-conversion photons are 809.7 nm (0.4 nm) and 1551 nm (1.5 nm). The down-conversion bandwidths were measured using difference-frequency generation~\cite{DFGJSI} between the pump and a tunable 1550-nm laser, whose wavelength was swept while the counts on the 810-nm side were recorded~\cite{Trentthesis}. The outputs of the Sagnac entangled photon source are coupled into single-mode fiber before going to the analyzers. Unintended wedge of the PPLN crystal produces lateral walk-off of the (counter-)clockwise paths which lowers the heralding efficiency to about 15\% or less; other Sagnac implementations have achieved efficiencies exceeding 80\%~\cite{Ramelow:13}. Moreover, other SPDC configurations can also be used to create high-fidelity, high-brightness, and/or high-heralding efficiency polarization-entangled photon-pair sources; these include non-colinear two-crystal sources~\cite{PhysRevLett.75.4337}, beam-displacer sources~\cite{BD1SPDC}, and waveguide sources~\cite{Takesue:08}.

Avalanche photodiodes (Excelitas SPCM-AQ4C) with about 45\% efficiency were used to detect the 810-nm photons. The 1550-nm photons were detected using four WSi superconducting nanowire detectors from NASA's Jet Propulsion Laboratory, optimized for 1550 nm with an efficiency of approximately 80\%~\cite{chapCLEO2017}. However, one detector had efficiency of only 40\%, so 3-dB attenuators were added to the fibers entering the other detectors to even out the detection efficiency for HEQKD measurements; since our goal was to simulate channels with much higher losses, this 3-dB factor can be interpreted as part of the link in a future implementation where all detectors have $>80$\% efficiency.

For BBM92, the measurement basis is randomly chosen by the 50/50 non-polarizing first beamsplitter in Alice or Bob's delay interferometer: If the photon goes through the short path, it is analyzed in the H/V basis; if it travels the long path, it is analyzed in the D/A basis because of a half-wave plate (HWP) that effectively rotates the basis of that beamsplitter port from H/V to D/A. For HEQKD, the measurements of the four bases are made using an analyzer interferometer, polarization optics, and a time-bin sorting circuit. The analyzer interferometer allows for measurements of superpositions of the time bins, while the polarization optics allow for measurements of different combinations of polarization and time bin. Measurements in bases 1 and 2 (or 3 and 4) are distinguished by a time-bin sorting circuit that is, electrically, in between the detectors and the time taggers; see Appendix \ref{tbs} for explanation of the operation of this circuit. See Table \ref{BBM92detmapping} (\ref{HEQKDdetmapping}) for a complete description of the mapping between the bases and measurements to the output time bins and detectors for BBM92 (HEQKD). 

In our system, the probability of basis 1 and basis 2 is split evenly, as is the probability between basis 3 and basis 4, because of the 50/50 non-polarizing beamsplitters in Alice's and Bob's analyzers. Additionally, we chose to measure in basis 1 and basis 2 $2p=50\%$ of the time (controlled by rotating a HWP at the front of the analyzer interferometer), to measure an even sampling of all bases for the QBER characterization to follow. Later in our analysis, however, we calculate the optimal value for this parameter as a function of channel loss and in a space-to-ground channel, assuming a fast electro-optic modulator is used to randomly choose the bases for each laser pulse. Due to our system's unique construction, we are able to make the full mutually unbiased basis measurement required for secure QKD with polarization and time-bin hyperentangled photons and use photons residing in all output time-bins. In contrast, the hyperentanglement in polarization and energy-time setup used by Ecker et al.~\cite{Eckerhyperentang} relies on a post-selection-free measurement system (no time-bin filtering needed~\cite{PhysRevA.54.R1}) which precludes them from making the full mutually unbiased-basis measurement for their state.
\section{Results}
\subsection{Implementation Comparison}
\begin{figure}
\centerline{\subfloat[]{{\includegraphics[scale=0.5]{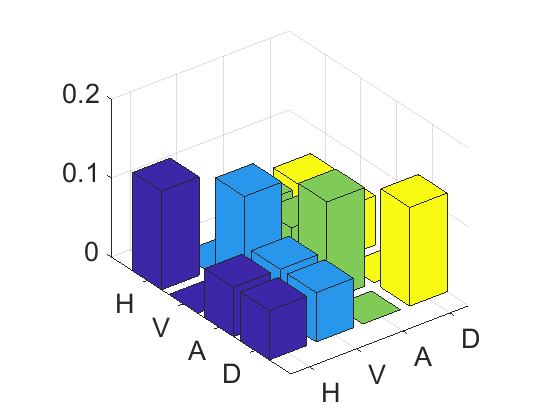}}}\subfloat[]{{\includegraphics[scale=0.5]{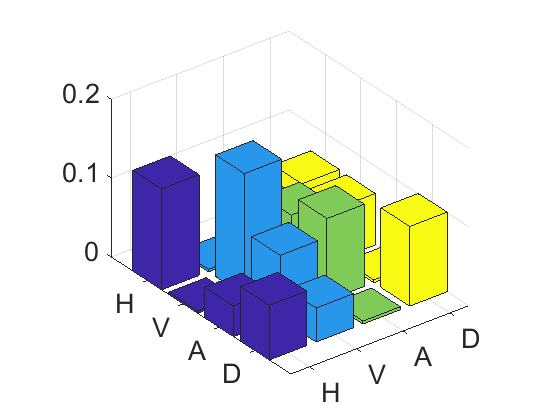}}}}\qquad\\\centerline{\subfloat[]{{\includegraphics[scale=0.7]{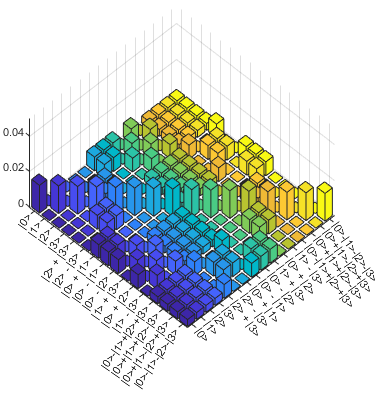}}}\subfloat[]{{\includegraphics[scale=0.7]{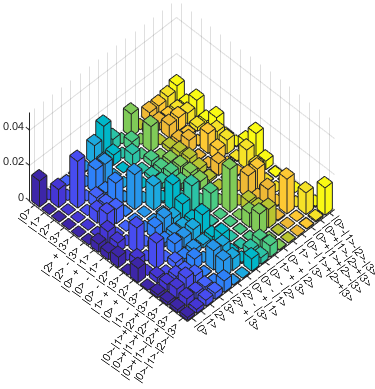}}}}\caption{{Crosstalk Matrices: The x-axis (y-axis) shows Bob's (Alice's) projective measurement. The whole crosstalk matrix is normalized by the sum of all elements. The QBER for a combination of bases is normalized by the sum of all elements in that basis combination. (a) For the ideal BBM92 normalized crosstalk matrix, H/V QBER $Q_{1,1,2,\text{est}}=0$ and D/A QBER $Q_{2,2,2,\text{est}}= 0$. (b) Measured BBM92 normalized crosstalk matrix, showing BBM92 QBER per basis for measured system (error in last digit calculated from the standard deviation assuming Poisson statistics): H/V QBER $Q_{1,1,2,\text{est}}= 0.0088(3)$ and D/A QBER $Q_{2,2,2,\text{est}}= 0.0185(6)$. (c) Ideal HEQKD normalized crosstalk matrix, where QBER ideally is zero in every basis combination along the diagonal and 50\% for all basis combination not along the diagonal or anti-diagonal, and $\ket{0}\equiv\ket{Ht_1^p}$, $\ket{1}\equiv\ket{Vt_2^p}$, $\ket{2}\equiv\ket{Vt_1^p}$, and $\ket{3}\equiv\ket{Ht_2^p}$. (d) Measured HEQKD normalized crosstalk matrix. The measured HEQKD QBERs per basis (error in last digit calculated from the standard deviation of seven measurement samples): $Q_{1,1,4,\text{est}}=0.010(3)$, $Q_{1,2,4,\text{est}}=0.45(1)$, $Q_{1,3,4,\text{est}}$ $Q_{2,1,4,\text{est}}=0.49(2)$, $Q_{2,2,4,\text{est}}=0.036(7)$, $Q_{2,4,4,\text{est}}=0.54(1)$, $Q_{3,1,4,\text{est}}=0.50(2)$, $Q_{3,3,\text{est}}=0.029(5)$, $Q_{3,4,\text{est}}=0.47(2)$, $Q_{4,2,\text{est}}=0.50(1)$, $Q_{4,3,\text{est}}=0.51(2)$, and $Q_{4,4,\text{est}}=0.051(9)$.}}\label{BBM92HEQKDgoodcrosstalk}
\end{figure}

Fig. \ref{BBM92HEQKDgoodcrosstalk} shows the basic performance of our BBM92 and HEQKD implementations~\footnote{There are two common forms to characterize the error of measured values, the standard deviation (a measure of the sample set's variability away from the sample mean) and the standard error (a measure of the variability of a sample mean away from the population mean)~\cite{altman2005standard}. Here, and throughout this paper, we chose to exclusively use standard deviation because that is the correct quantity to calculate when trying to quote the spread of a given sample set. Additionally, we wanted to use standard deviation, so the systematic error seen from environmental fluctuation and time-dependent calibration errors were accounted for, which is not the case for the standard error, which assumes the population mean is constant, i.e., no environmental fluctuation and time-dependent calibration errors. Furthermore, in QKD often the goal is to bound the chance the QBER is larger than a certain value, so standard deviation is the appropriate quantity.}. The ``crosstalk matrix'' shows the expected correlations and anti-correlations when the same measurement basis is used by Alice and Bob and the partially correlated or uncorrelated results when they use different bases. Experimentally, we see QBER below 2\% for BBM92 and below 5.5\% for HEQKD (for bases that are perfectly correlated). There are some imbalances in the matrix element probabilities between different bases and within bases due to transmission variability for different measurements and bases.

Due to the non-uniform amount of correlation between different basis combinations in HEQKD, there are some basis combinations that have higher QBER, which is expected. For the basis combinations where Alice and Bob measure in the same basis, we ideally expect perfect correlations, i.e., along the diagonal, and therefore, low QBER in practice and up to 2 bits of raw key to be extracted per detection event. For the other basis combinations, there is half the correlation, so we ideally expect about 50\% QBER. For example, in Fig. \ref{BBM92HEQKDgoodcrosstalk}c, in the sub-crosstalk matrix between basis  $\mathsf{A}_1$ and $\mathsf{B}_3$, the 4 diagonal probabilities are ``correct" outcomes since there is correlation between Alice and Bob; anything else off the diagonal is considered ``incorrect". In that sub-crosstalk matrix, half of the probability is along the diagonal and half is not, resulting in a QBER of 50\%. For $D=4$, there are 4 possible outcomes; since only one of them is the correct one, 3/4 of the time there will be an error. In general, when there is 1 correct outcome and d-1 incorrect outcomes, the error probability of random uncorrelated events is $e_0=(d-1)/d$. Therefore, in a $D=4$ protocol, random events give a 75\% QBER, so even at 50\% QBER, there is still up to one bit of raw key to be extracted. Note this means that not only does higher-dimensional encoding yield more efficient secret key capacity (secret bits per photon detected), it can also allow operation in regimes of higher QBER. For example, where BB84 and BBM92 can tolerate at most QBER = 11\% in the assymptotic limit, before secure key fraction vanishes, whereas a $D=4$ protocol permits QBERs as high as 18.93\%~\cite{NDQKD}.
\captionsetup[subfigure]{position=top, singlelinecheck=false,justification=raggedright,captionskip=0pt}

\begin{figure}
\centerline{\subfloat[]{{\includegraphics[scale=0.5]{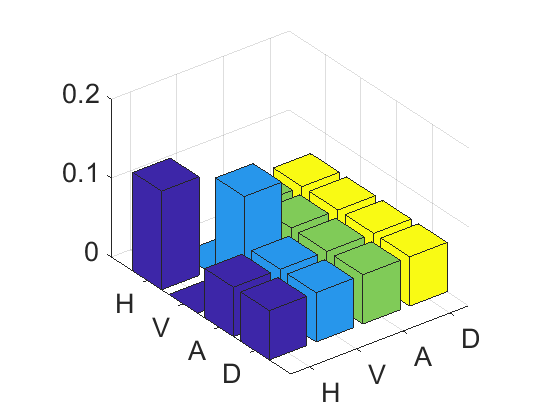}}}\subfloat[]{{\includegraphics[scale=0.5]{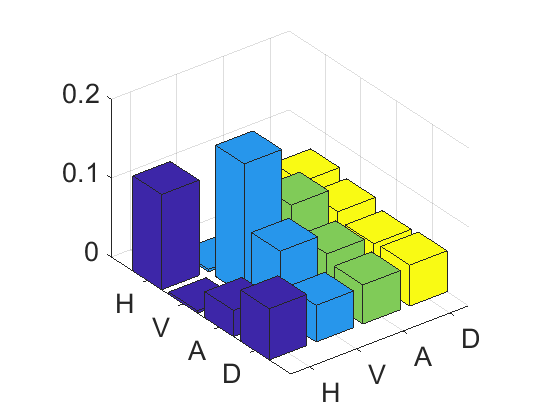}}}}\qquad\\
\centerline{\subfloat[]{{\includegraphics[scale=0.65]{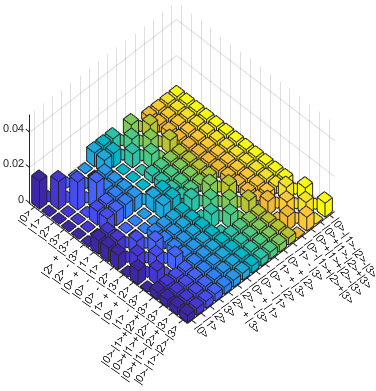}}}\subfloat[]{{\includegraphics[scale=0.65]{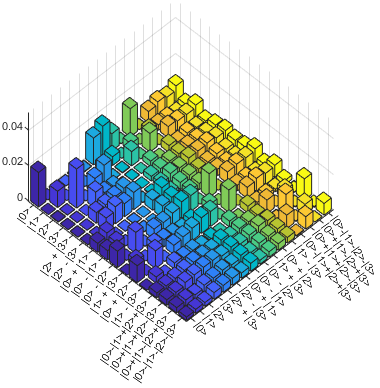}}}}
\caption{{Crosstalk Matrices With Eavesdropper: The x-axis (y-axis) shows Bob's (Alice's) projective measurement. The whole crosstalk matrix is normalized by the sum of all elements. The QBER for a combination of bases is normalized by the sum of all elements in that basis combination. (a) Ideal BBM92 normalized crosstalk matrix when an eavesdropper performs an ideal intercept-resend attack on the H/V basis, yielding H/V QBER $Q_{1,1,2,\text{est}}=0$ and D/A QBER $Q_{2,2,2,\text{est}}= 0.5$. (b) Measured BBM92 normalized crosstalk matrix with a birefringent crystal to decohere the polarization in the H/V basis, yielding measured [ideal] H/V QBER $Q_{1,1,2,\text{est}}= 0.028[0.00]$ and D/A QBER $Q_{2,2,2,\text{est}}= 0.48[0.50]$. (c) Ideal HEQKD normalized crosstalk matrix when an eavesdropper performs an ideal intercept-resend attack on the polarization qubit, where $\ket{0}\equiv\ket{Ht_1^p}$, $\ket{1}\equiv\ket{Vt_2^p}$, $\ket{2}\equiv\ket{Vt_1^p}$, and $\ket{3}\equiv\ket{Ht_2^p}$. (d) Measured HEQKD normalized crosstalk matrix with a birefringent crystal to decohere the polarization in the H/V basis. QBER per basis for measured [ideal] system: $Q_{1,1,4,\text{est}}=0.02[0.00]$, $Q_{1,2,4,\text{est}}=0.47[0.50]$, $Q_{1,3,4,\text{est}}=0.53[0.50]$, $Q_{2,1,4,\text{est}}=0.52[0.50]$, $Q_{2,2,4,\text{est}}=0.51[0.50]$, $Q_{2,4,4,\text{est}}=0.74[0.75]$, $Q_{3,1,4,\text{est}}=0.53[0.50]$, $Q_{3,3,4,\text{est}}=0.50[0.50]$, $Q_{3,4,4,\text{est}}=0.74[0.75]$, $Q_{4,2,4,\text{est}}=0.73[0.75]$, $Q_{4,3,4,\text{est}}=0.74[0.75]$, and $Q_{4,4,4,\text{est}}=0.53[0.50]$.}}\label{BBM92HEQKDevecrosstalk}
\end{figure}

To verify the ability of BBM92 and HEQKD to detect an eavesdropper, we inserted, in the channel to Bob, a 9-mm-thick a-cut calcite crystal, oriented so that H and V polarizations are relatively unaltered but travel at different speeds. The calcite is thick enough so the H and V polarizations can no longer interfere after exiting the calcite, because their wavepackets no longer overlap temporally. Specifically, the calcite introduces about a $5$-ps relative delay, much larger than the $1.7$-ps coherence time of the 1.5-nm (0.4-nm) bandwidth 1550-nm (810-nm) photons. This simulates an eavesdropper that measures only in the H/V basis and simply records, then resends to Bob, what is measured~\footnote{Note this implementation also models an eavesdropper who entangles the photon to a quantum processor for later analysis. Such coupling must necessarily occur in some basis; tracing over the states of the eavesdropper's processor will then remove the coherence of the intercepted photon in the coupling basis.}. With this technique, the eavesdropper gains a significant amount of information at the necessary cost of introducing a significant amount of errors. Fig. \ref{BBM92HEQKDevecrosstalk} shows the expected and measured results of the eavesdropping. As expected, the results show a greatly increased QBER in all bases involving measurements using superpositions of polarizations (and only a slight increase in the H/V basis QBER due to imperfect calcite alignment).

\begin{figure}
\centerline{\subfloat[]{{\includegraphics[scale=0.4]{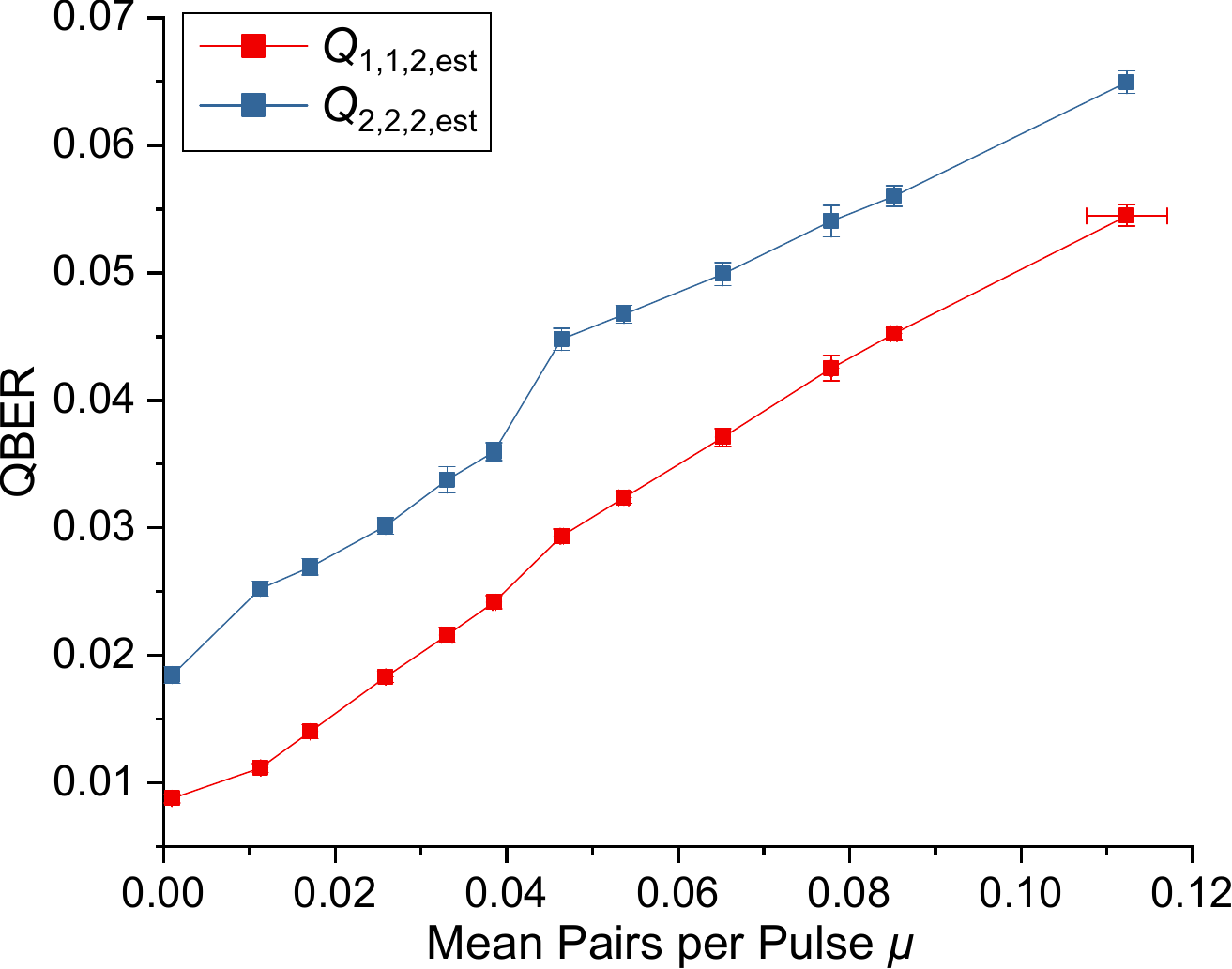}}}\subfloat[]{{\includegraphics[scale=0.4]{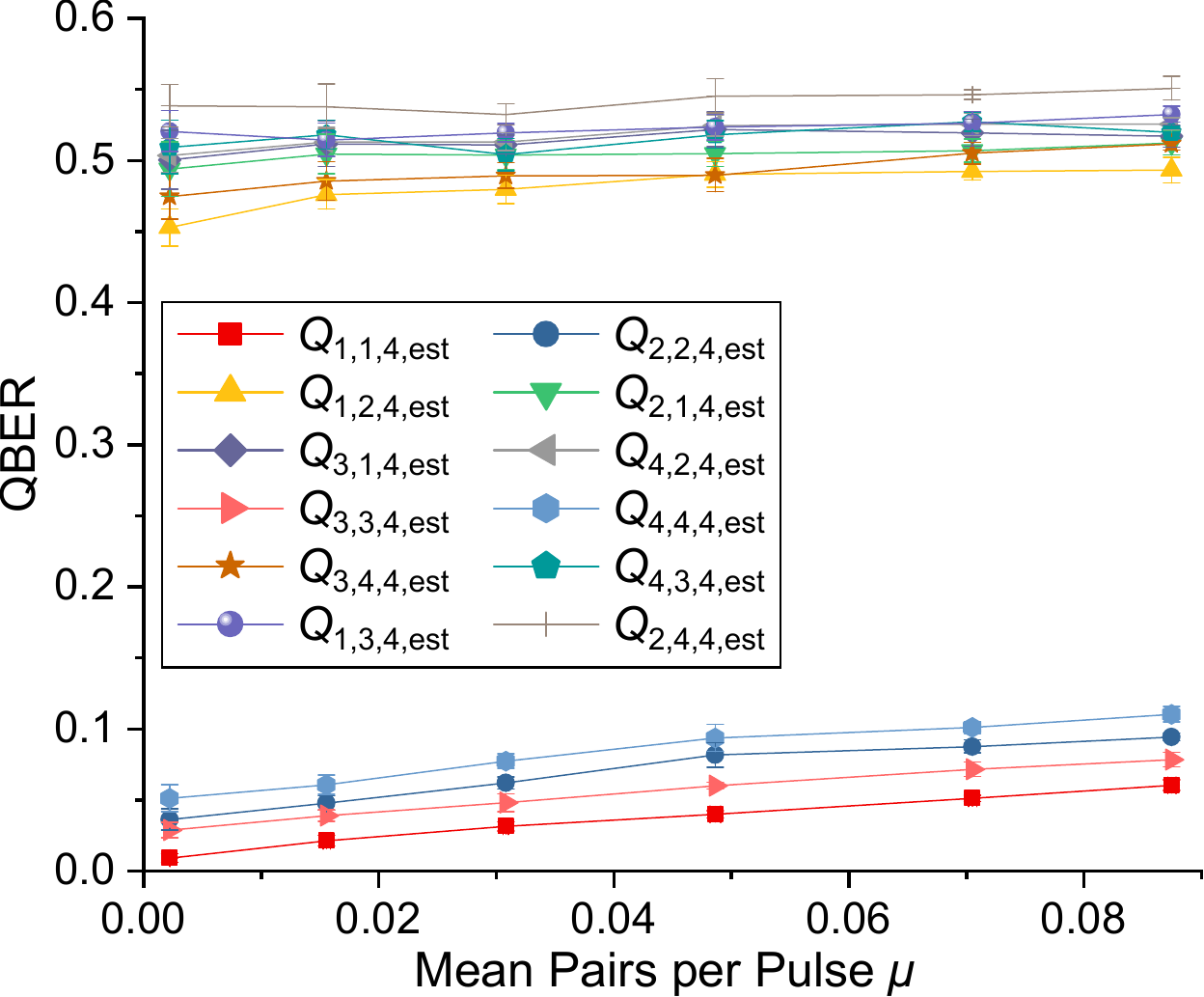}}}}
\caption{Measured QBER versus mean pairs per pulse, adjusted by varying the pump-laser power, for each key-generating basis combination: (a) BBM92 results, when the total losses for Alice and Bob are 12 dB and 15 dB, respectively. The error bars, created from ten measurements, correspond to one standard deviation. (b) HEQKD results, when the total losses for Alice and Bob are 15 dB and 18 dB, respectively. The error bars, here determined from seven measurements, correspond to one standard deviation.}\label{BBM92HEQKDmu}
\end{figure}

The intrinsic QBER of the system can be measured accurately for all bases when the probability of producing an entangled pair is sufficiently low so that the probability of producing multiple pairs (which can cause errors) is negligible. But since this also reduces the key rate, there is a trade-off, leading to an optimal pump power (mean pairs per pulse, $\mu$) that produces the most secret key per use, especially when accounting for finite statistics. Fig. \ref{BBM92HEQKDmu} shows the measured QBER versus mean pairs per pulse. For each data set, the mean pairs per pulse were calculated from solving this system of equations:
\be
S_i=\frac{RT(\eta_i\mu(4+\eta_i\mu)+4\xi_i)}{(2+\eta_i\mu)^2}\text{,}
\ee
\be
C_{AB}=RT\Big( 1- \frac{4(1-\xi_A)}{(2+\eta_A\mu)^2}-\frac{4(1-\xi_B)}{(2+\eta_B\mu)^2}+\frac{4(1-\xi_A)(1-\xi_B)}{(2+\eta_A\mu + \eta_B\mu-\eta_A\eta_B\mu)^2}\Big)
\text{,}
\ee
where $S_i$ is the singles rates for Alice or Bob, $C_{AB}$ is the total coincidence rate between Alice and Bob, $R$ is the repetition rate of the laser, $T$ is the integration time, $\eta_i$ is the transmission for Alice or Bob, and $\xi_i$ is the noise counts per laser pulse for Alice or Bob~\cite{Ma2007}. 

The data shows a significant variation in the intrinsic QBER of each basis (as $\mu\rightarrow0$), originating from the different physical processes that are present in the states and projections measured in each basis. The QBER in the H/V basis of BBM92, and basis 1 of HEQKD, is only affected by imperfect H/V basis alignment between Alice and Bob and imperfect polarizing beamsplitter extinction ratio, leading to only about $1\%$ QBER ($\ket{HV}$-type terms in our state generation could also cause this same QBER, but are apparently negligible, since the basis 1 QBER matches our classical measurements of the basis alignment and polarizing beamsplitter). HEQKD basis 2 is affected by the same influences as basis 1 but is also affected by the temporal entanglement visibility (about $96\%\rightarrow2\%$ QBER) and imbalances in the measured amplitudes of the terms in the superposition (about $0.5\%$ QBER). The QBER in the D/A basis of BBM92, and HEQKD basis 3, is also affected by the same error processes as basis 1 and is additionally affected by imperfect polarization entanglement purity of the source (D/A visibility about $98\%\rightarrow1\%$ QBER) and imbalances in the measured amplitudes of the terms in the superposition (about $0.5\%$ QBER). Finally, HEQKD basis 4 is affected by all previously mentioned error processes, and thus has the highest total QBER (about 5\%).

\begin{figure}
\centerline{\subfloat[]{{\includegraphics[scale=0.405]{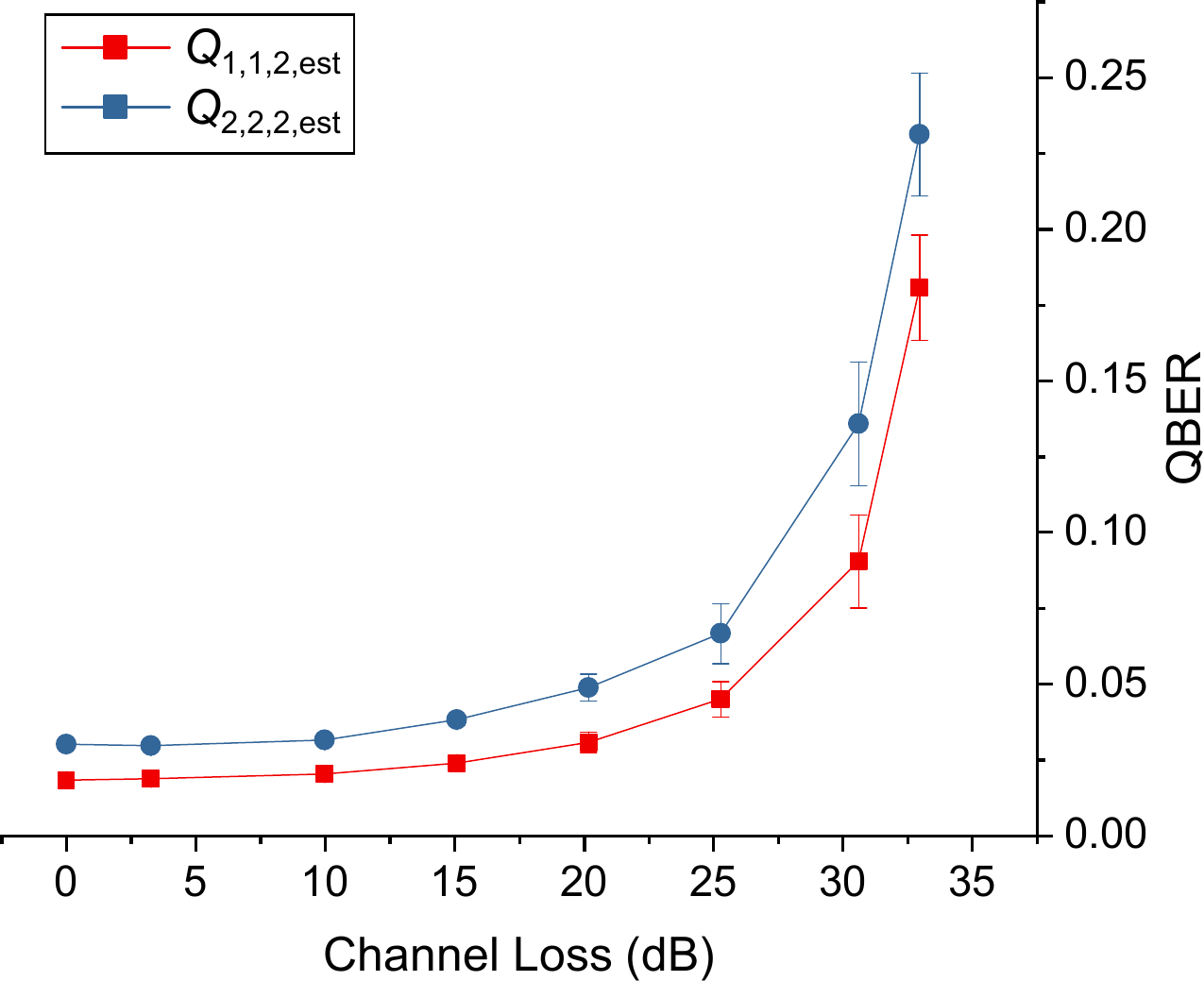}}}\subfloat[]{{\includegraphics[scale=0.4]{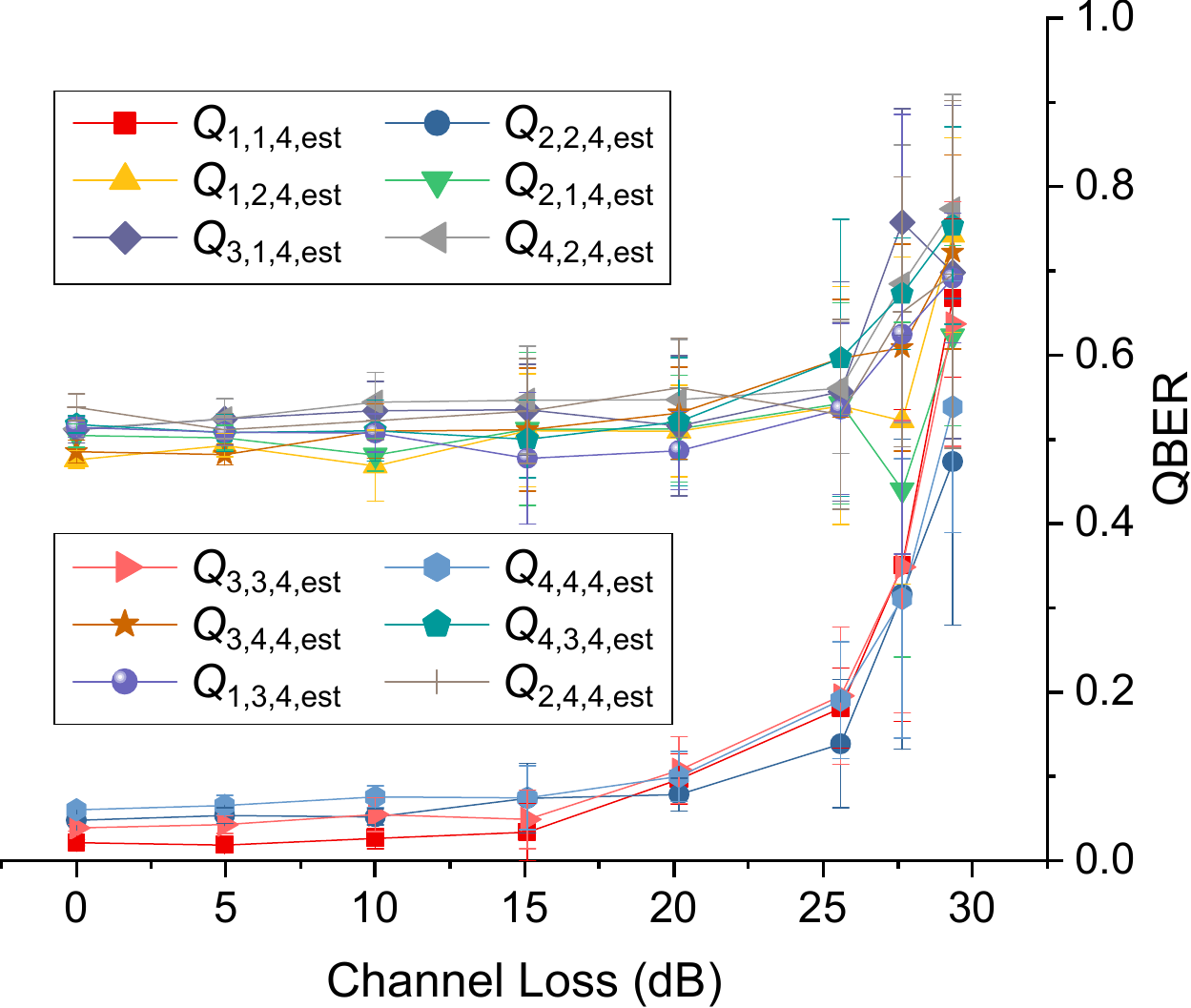}}}}
\caption{{QBER versus Channel loss between the entangled photon source and Bob, for each key-generating basis combination: (a) BBM92 results, with Alice's total loss fixed at 12 dB; Bob's minimum intrinsic loss, without any channel loss, is 15 dB, and the mean pairs per pulse was $\mu=0.026$. The error bars, determined from at least ten measurements (up to 55 taken for higher loss data points), correspond to one standard deviation. (b) HEQKD results, with Alice's total loss now fixed at 15 dB; Bob's minimum intrinsic loss is 18 dB, and the mean pairs per pulse was $\mu=0.016$. The error bars, determined from eight measurements, correspond to one standard deviation.}}\label{BBM92HEQKDloss}
\end{figure}

Since there would be a finite time-window during orbit when one could establish the required line-of-sight quantum channel (see link analysis below), and the channel transmission will change during that window as the range changes, it was important to characterize the system for various values of channel transmission. Errors from detector noise and background events will start to dominate the data when the channel transmission is too low. Fig. \ref{BBM92HEQKDloss} shows the measured QBER for each basis in each protocol, for decreasing values of Bob's channel transmission. At the highest measured transmission in Fig. \ref{BBM92HEQKDloss}, our system (which was optimized for QBER not key rate) produced about 2000 sifted events/s for BBM92 and about 1000 sifted events/s for HEQKD, correcting for the different $\mu$ used in each protocol for this measurement (see Fig. \ref{BBM92HEQKDloss}). At low channel loss, some bases start at about 50\% QBER because those bases are not fully correlated. Since there are four measurement outcomes per basis, the 75\% QBER expected  for a completely random input implies there is still some information to be extracted from those bases at 50\% QBER. But, as expected, when the channel loss increases, the QBER increases to near 75\% for all bases, as the raw key is increasingly created from random noise events.

\subsection{Doppler Shift}
\label{sec:Dopp}
\begin{figure}
\centerline{\subfloat[]{{\includegraphics[scale=0.4]{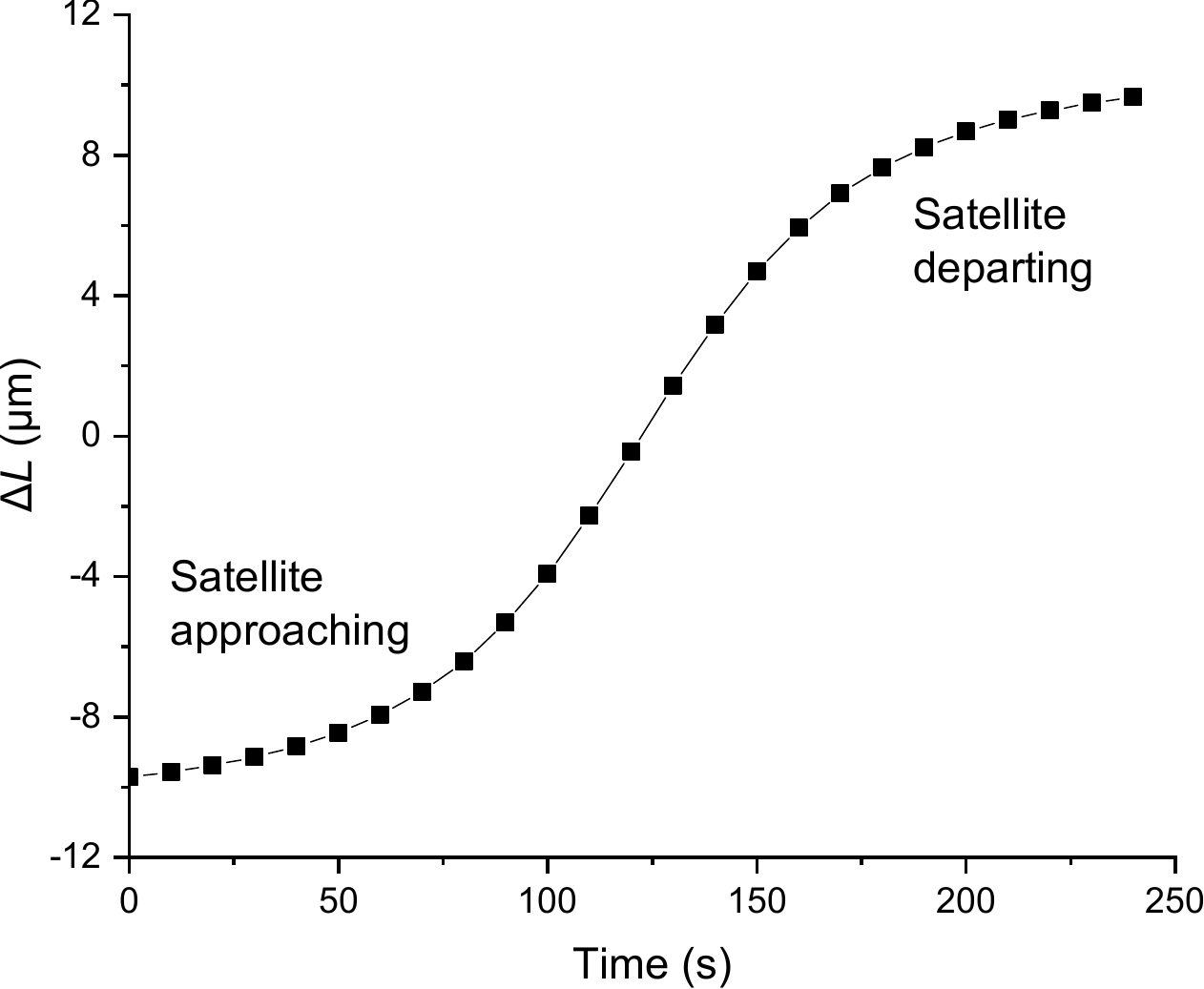}}}\subfloat[]{{\includegraphics[scale=0.35]{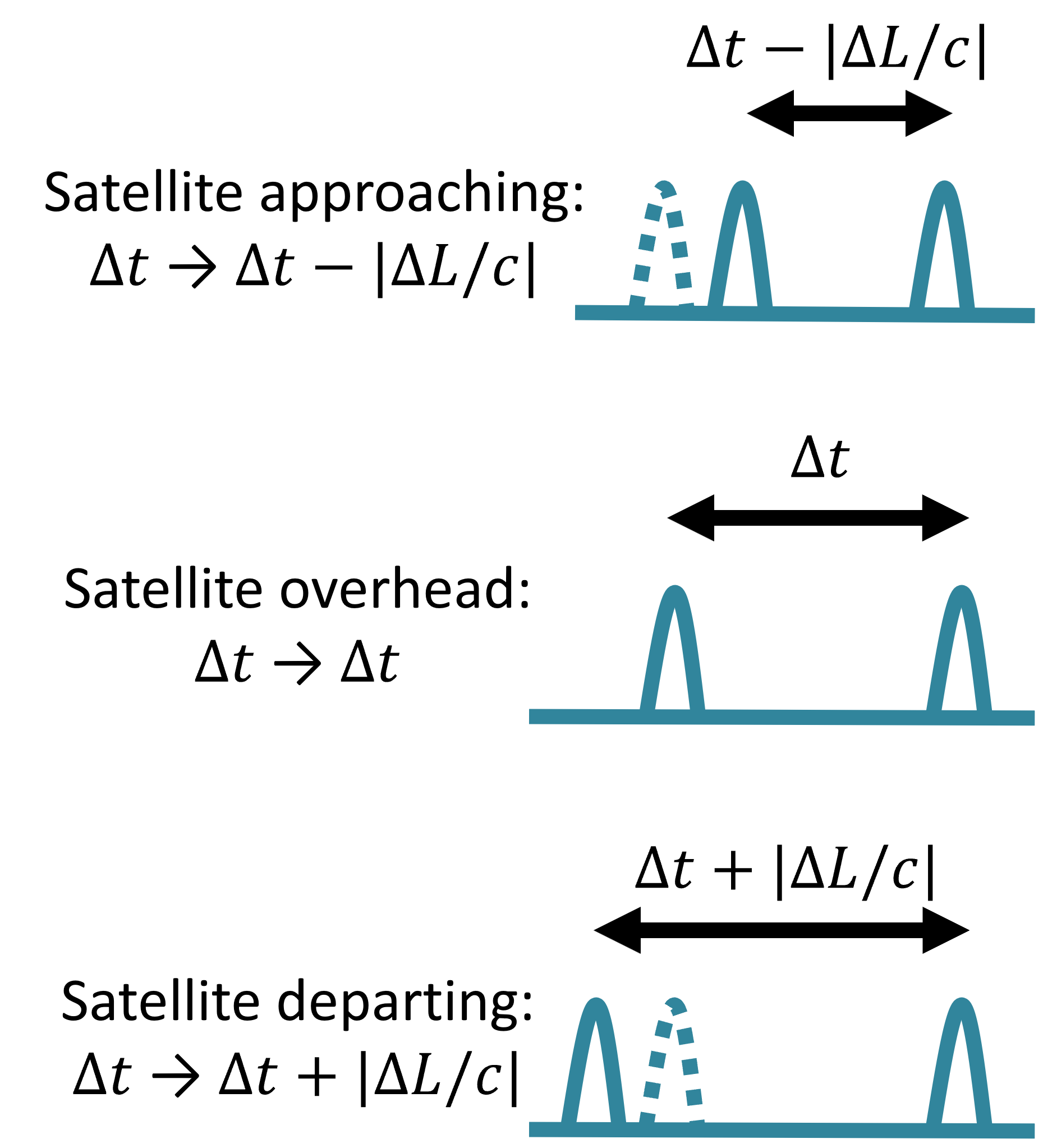}}}}\caption{Expected Doppler Shift: (a) Expected Doppler shift for an overhead orbit. (b) Pictorial explanation of effect of Doppler shift on time bins.}\label{Doppfigexplan}
\end{figure}

Since any transmitter in space will be moving rapidly, the interval between adjacent time bins in the ground station's reference frame initially will be reduced with respect to the interval measured in the transmitter's reference frame, as the transmitter is approaching. The intervals will match as the transmitter passes overhead, and as it moves away, the received interval will be longer (Fig. \ref{Doppfigexplan}b)~\footnote{There is also a frequency shift on the photons, but since $\gamma\equiv\frac{1}{\sqrt{1-(\frac{V_{sat,long}}{c})^2}}=1.00000000033$, assuming $V_{sat,long}=7.7$ km/s (average velocity, e.g., of the ISS), the wavelength shift is negligible ($\delta\lambda\approx\lambda(1-\gamma)$) compared to the photon bandwidth of about $1$ nm.}.

Assuming there is an adaptive optics system on the ground station to correct for turbulence (effectively a transverse variation in phase), the Doppler shift is the only source of time-bin phase change from the space-to-ground channel~\footnote{The effect of changing ``piston'' due to random density fluctuations along the path is completely negligible given the short interval between the time bins.}. The exact phase shift produced through the Doppler effect is dependent on several parameters, including the maximum elevation angle of the orbit during a pass, which changes for subsequent passes and is at a maximum for passes directly overhead, i.e., when the maximum elevation angle is $90^{\circ}$. For time bins separated by 1.5 ns and an overhead pass, calculations of the relativistic longitudinal Doppler shift~\cite{reldop} using a simulation of a LEO (low-Earth orbit) satellite with orbit inclination of $51^{\circ}$ (angle between orbital plane and equator), 400-km altitude, i.e., the orbit of the International Space Station, and with the longitudinal velocity, $V_{sat,long}$, calculated along the beam path with respect to a ground station at $39^{\circ}$ latitude (e.g., continental United States), show an expected shift of 
\begin{equation}
\Delta{L}(t)=\Bigg[\sqrt{\frac{1+\frac{V_{sat,long}(t)}{c}}{1-\frac{V_{sat,long}(t)}{c}}}-1\Bigg](1.5\text{x}10^{-9}s)c\text{,}
\end{equation}
as displayed in Fig. \ref{Doppfigexplan}a. If acquisition starts and stops at a $20^{\circ}$ elevation angle (below which atmospheric scattering and turbulence make relatively efficient single-photon collection infeasible), then the total change in pulse separation during a pass of the satellite is $\Delta{L}(t_{stop})-\Delta{L}(t_{start})=12.8$ $\mu$m.

\begin{figure}
\centerline{\subfloat[]{{\includegraphics[scale=0.4]{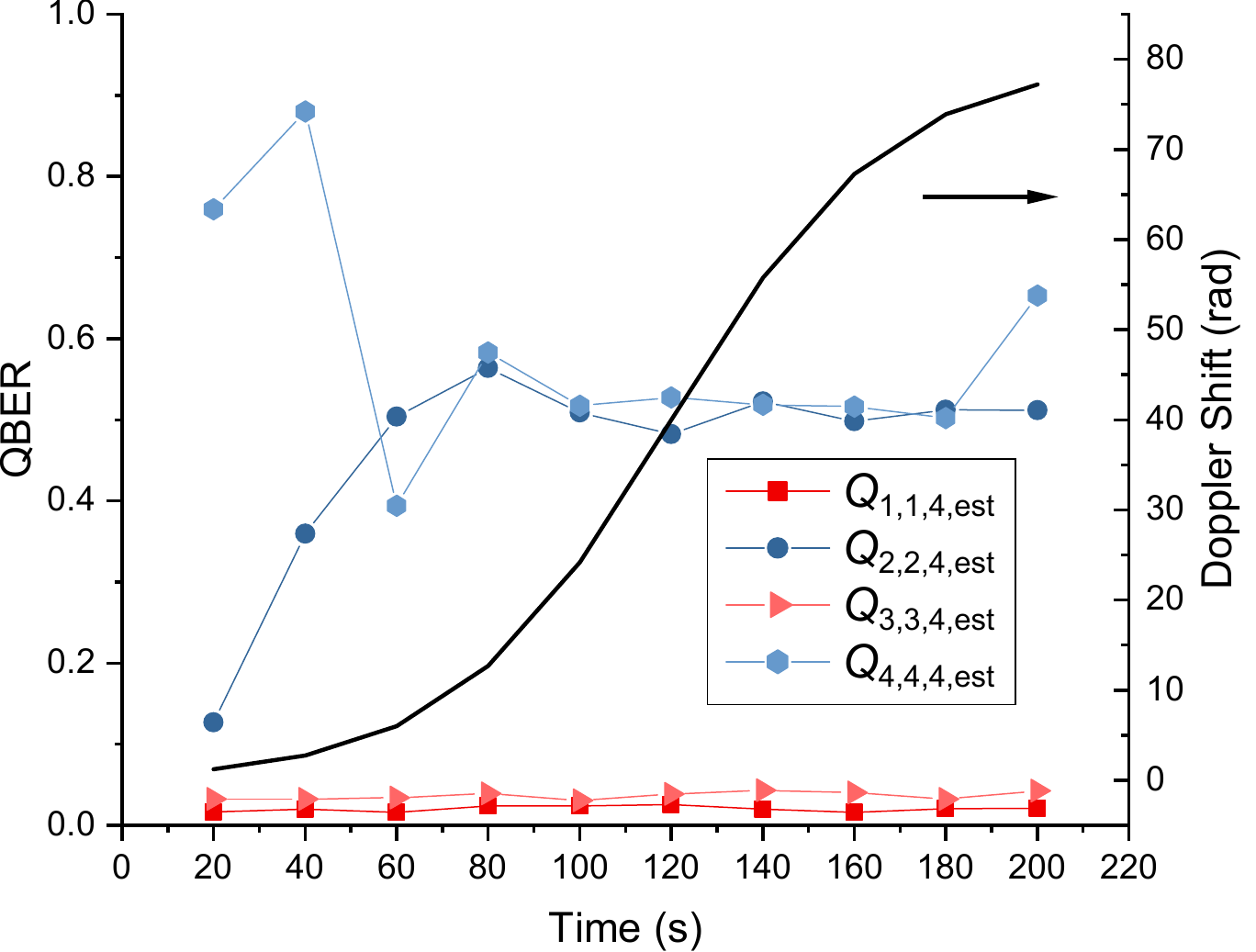}}}\subfloat[]{{\includegraphics[scale=0.4]{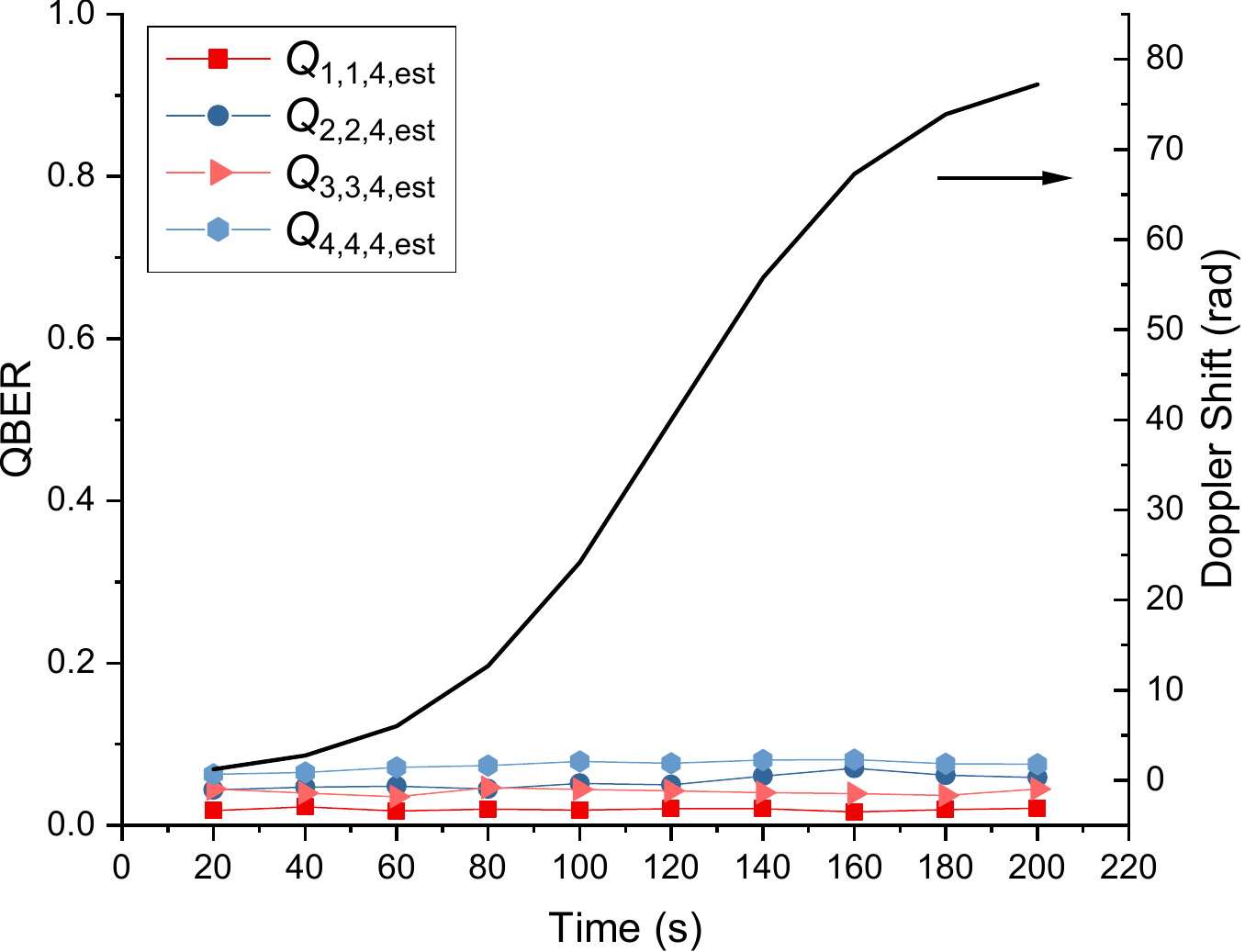}}}}\caption{Doppler Shift Effect on HEQKD: Measured QBER for matching basis combinations during an in-lab emulated Doppler shift (solid black curve) (a) without phase stabilization; (b) with phase stabilization active.}\label{HEQKDdoppler}
\end{figure}

We implemented an in-lab emulation of this Doppler shift by smoothly moving a piezo-actuated translation stage that controlled the position of the pump's right-angle prism, using the same distance-versus-time profile as in Fig. \ref{Doppfigexplan}a. To keep this Doppler shift (and any other time-varying phase shifts, e.g., from local vibrations on the transmitting satellite) from adversely affecting the protocol's performance, we developed a phase-stabilization system that uses a classical laser beam and proportional-integral feedback to track the path-length difference of the ground interferometer, so it matches that of the emitted time bins throughout the pass (see Appendix \ref{stab} for more information). As seen in Fig. \ref{HEQKDdoppler}a, QKD using time bins is not possible without such phase stabilization because the QBER in some bases is too high, even though other bases (specifically, bases 1 and 3) are unaffected by the Doppler shift (since their basis states, e.g., polarization, are time-bin phase insensitive). QBER in some bases, e.g., Alice and Bob both chose basis 4, starts high at Time = 0 s because there was no initial time-bin phase calibration done for this dataset. Fig. \ref{HEQKDdoppler}b shows the performance of HEQKD while a lab-simulated Doppler shift was occurring, with initial phase calibration done and the phase stabilization activated. 

During the Doppler shift the QBER was held stable to a standard deviation $<1\%$. Since the quantum signal and classical-phase-stabilization pulses co-propagate with one another, we expect nearly all the potential phase drift caused by satellite motion and vibration to be common-mode except for phase drifts in the ground station analyzer. Therefore, we would expect similar performance of our lab-based phase stabilization system in an actual space-to-Earth implementation.
\subsection{Secret-Key-Rate Simulations}\label{SKRS}
To perform a full finite-key optimization for the maximum secret key rate, we simultaneously optimized, via sequential quadratic programming~\cite{nocedal2006numerical}, the mean pairs per pulse, $\mu$ (see Fig. \ref{fksim}a), the HEQKD basis probability, $p$, and the BBM92 parameter estimation ratio, $r$ (see Fig. \ref{fksim}b), truncating the infinite sums that implicitly contribute in Eqn. \ref{qberest} at $n=10$ photon pairs per pulse (see Appendix \ref{FKA} for details). Since $P_{10}$ pairs is about $10^{-10}$ for $\gamma=0.1$, this truncation negligibly affects our simulations, but makes the simulations significantly easier to calculate and optimize.

Because the QBERs~\footnote{For these simulations, we extracted the average instrinsic QBER for each basis combination by applying a linear fit to each curve in Fig. \ref{BBM92HEQKDmu} and then using the y intercept.} in different bases vary significantly for our HEQKD implementation (c.f. Fig. \ref{BBM92HEQKDmu}b) it is not optimum to use a balanced analysis protocol (i.e., measure in all bases with equal likelihood); instead, in the asymptotic key regime (corresponding to losses of 0-35 dB), one should give preference to bases used for key generation, while in the finite-key regime ($>35$ dB) one should give more preference to bases used for error-checking, so that the error estimate is tighter in the finite key analysis (see Fig. \ref{fksim}b). Similarly, the optimal parameter estimation ratio for BBM92 is high in the asymptotic key regime and decreases in the finite-key regime, since more key is needed for accurate parameter estimation when the raw key length is shorter.
\subsubsection{Geo-stationary Orbit (GEO) Link Analysis}
\begin{figure}
\centerline{\subfloat[]{{\includegraphics[scale=0.4]{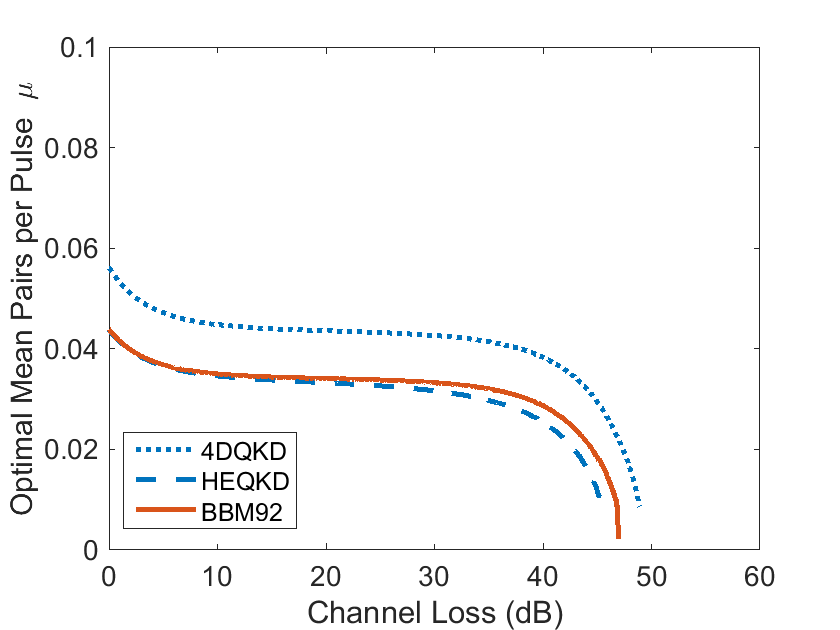}}}\subfloat[]{{\includegraphics[scale=0.4]{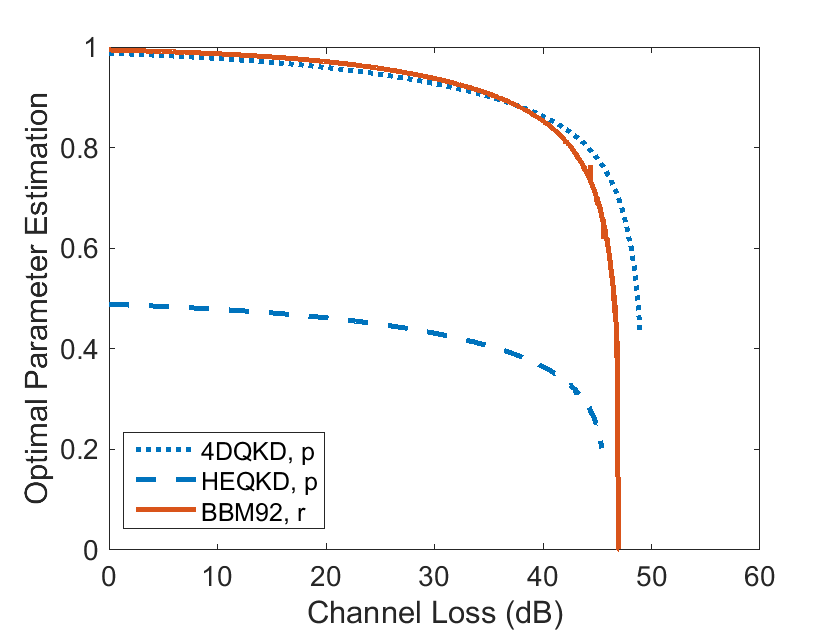}}}}\qquad\\
\centerline{\subfloat[]{{\includegraphics[scale=0.4]{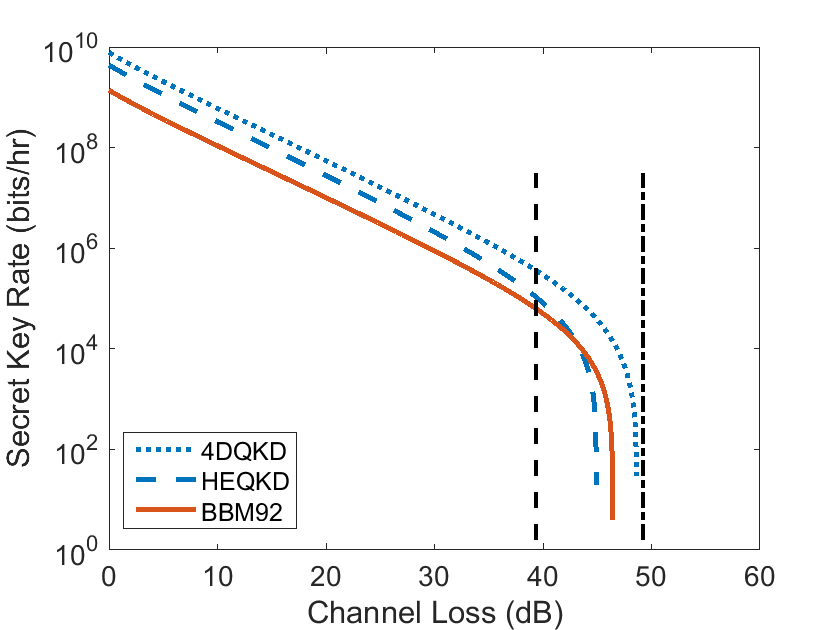}}}}\caption{Secret Key Length Optimization for HEQKD and BBM92: (a) Optimal mean pairs per pulse $\mu$ versus Bob's channel loss, for one hour of accumulated key generation. (b) Optimal 4DQKD basis probability $p$, optimal HEQKD basis probability $p$, and optimal BBM92 parameter estimation ratio $r$, versus channel loss, for one hour of key generation. (c) Estimated secret key rate (bits/hr) versus Bob's channel loss, calculated using the optimized values for $\mu$ and $p$ and feasible future system parameters: 400-MHz laser repetition rate, $10^{-6}$ background noise, and Alice's total efficiency including detection efficiency = 0.3. See Table \ref{sysparam} for more details. Assuming a 50-cm transmitter-aperture diameter and a GEO-altitude of 35,800 km, the dashed (dashed-dotted) lines correspond to the channel loss 39 dB (49 dB) for a 3-m (1-m) receiver-aperture diameter on the ground (see Secret-Key-Rate Simulations, Sec. \ref{SKRS}).}\label{fksim}
\end{figure}

Optimizing to obtain the longest final key for a long-distance HEQKD implementation, e.g., corresponding to a GEO link, with losses exceeding 40 dB, we find the optimal $\mu$ in Fig. \ref{fksim}a is nearly identical to that for BBM92. Since HEQKD encodes more raw-key per photon than BBM92, this leads to a higher secure key rate, as shown in Fig. \ref{fksim}c. For further comparison, we model a more standard $D=4$ QKD, which we call 4DQKD, using only 2 mutually unbiased bases~\footnote{Additionally, the definition of the basis probability is changed to $1=p+q$, where $p$ is the probability of choosing basis 1 and $q$ is the probability of choosing basis 4.}, in this case Bases 1 and 4. This shows that the HEQKD protocol introduced in Sec. \ref{sec:ProtIntro} is a type of hybrid between BBM92 and 4DQKD. We did not implement 4DQKD in our experiment for technical reasons~\footnote{It is not possible to build an analyzer for 4DQKD with passive linear optics without also measuring Bases 2 and 3. To measure just Bases 1 and 4 would require a polarization-independent active switch, as has been demonstrated~\cite{Trentthesis,10.1117/12.2537081}.}; we just modeled it for comparison. The results in Fig. \ref{fksim} assume QBERs in all protocols equivalent to what we measured with our setup. However, we also found that further improvement in QBER resulted in about $20\%$ increases in secret key rate and negligible increases ($<1$ dB) in maximum channel loss with positive secret key rate, as the signal becomes dominated by noise in these high-loss regimes.

Notably, Fig. \ref{fksim}c shows that HEQKD can still yield useful secret keys (about $200$ kb after 1-hr integration) in an overhead geo-stationary orbit (GEO), with the secure key rate increasing substantially for longer integration times (about $10$ Mb for 1-day integration) due to finite statistics effects. Assuming 0.2 dB/km of loss in standard optical fiber, Fig.~\ref{fksim}c shows appreciable finite secure key for distances exceeding 200 km, which is significant for an entanglement-based QKD demonstration in fiber. Note that, BBM92 still has an appreciable amount of key at GEO-like channel losses too. Moreover, we observe that at higher losses, BBM92 can give some key even when HEQKD cannot; this somewhat increased number of detectors in HEQKD that produce a greater total number of background events (including dark counts), which dominate the true coincidence detection events at high channel loss. The left vertical dashed line in Fig. \ref{fksim}c is located at about 39 dB, calculated using the Friis equation to estimate channel transmission $\eta$ as a function of range $r$: $\eta(r)=(\pi{D_T}D_R/(4\lambda{r}))^2$~\cite{Friis,OCRD}, and $r=$35,800 km (for GEO), transmitting telescope diameter $D_T=0.5$ m, receiving telescope diameter $D_R=3$ m, and wavelength $\lambda=1550$ nm, with the added assumptions of a 6-dB loss for receiver telescope transmission~\cite{biswas2014optical} and the adaptive optics single-mode fiber collection efficiency~\cite{chen2015experimental}, negligible atmospheric loss, negligible loss from pointing, acquisition, and tracking (PAT) errors, and using the other system parameters listed in Table \ref{sysparam}. Similarly, the right vertical dotted line at 49 dB was calculated using the Friis equation with a receiving telescope diameter $D_R=1$ m. We note that in this case there is no secure key after 1-hr integration for any of the protocols considered here, due to finite statistics and background events. However,  longer integration times (on the order of 1 day) or lower background event rate ($<10^{-8}$ per pump pulse) could permit some secure key extraction.

\begin{table}
\caption{System Parameters: Current and expected future system parameters relevant to simulation.}
%\centerline{\includegraphics[scale=0.28]{sysparameterstablev8.png}}
\label{sysparam}
\begin{center}
	\begin{tabular}{ c| c c}
	\hline
	\hline
		  Parameters & Current System & Simulated System \\ \hline

		Laser Repetition Rate&80 MHz&400 MHz\\
		Detector Efficiency A&0.45&0.6\\
		Detector Efficiency B&0.4&0.9\\
		Measurement System Transmission&0.6&0.7\\
		Heralding Efficiency&0.15&0.7\\
		Background Noise per Pump Pulse&$10^{-5}$&$10^{-6}$\\
		Secrecy $\eps_\tn{sec}$&-&$10^{-9}$\\
		Correctness $\eps_\tn{cor}$&-&$10^{-12}$\\
	\hline
	\hline		
			\end{tabular}
\end{center}
\end{table}

\subsubsection{Low-Earth Orbit (LEO) Link Analysis}
\begin{figure}
\centerline{\subfloat[]{{\includegraphics[scale=0.4]{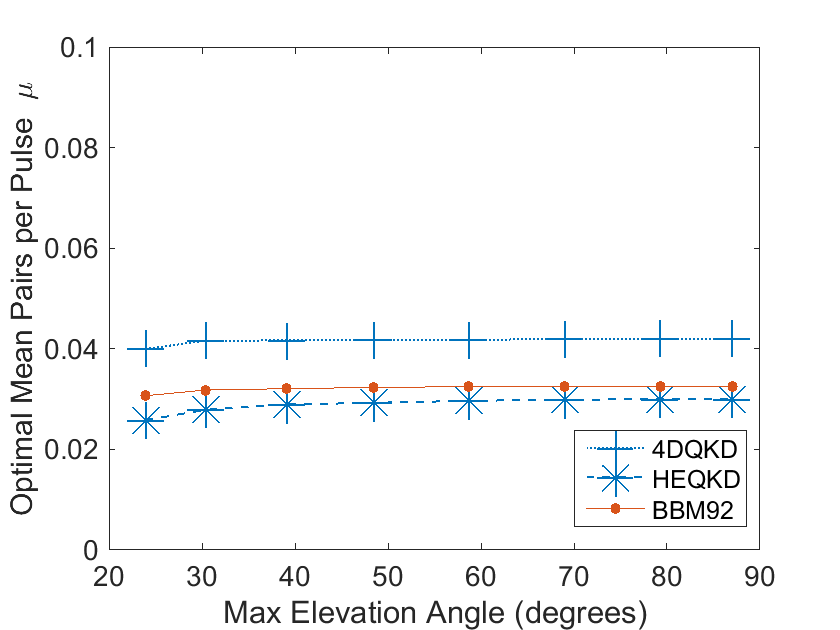}}}\subfloat[]{{\includegraphics[scale=0.4]{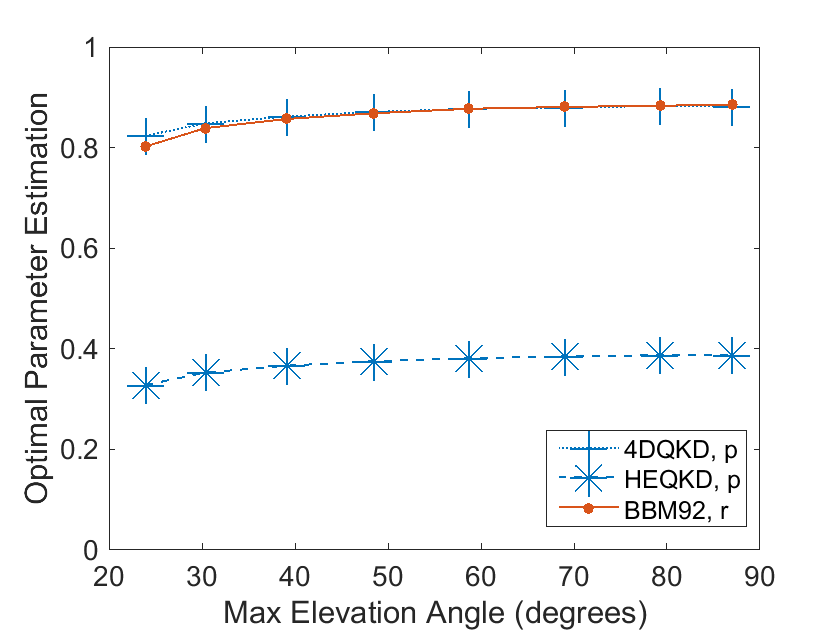}}}}\qquad\\
\centerline{\subfloat[]{{\includegraphics[scale=0.4]{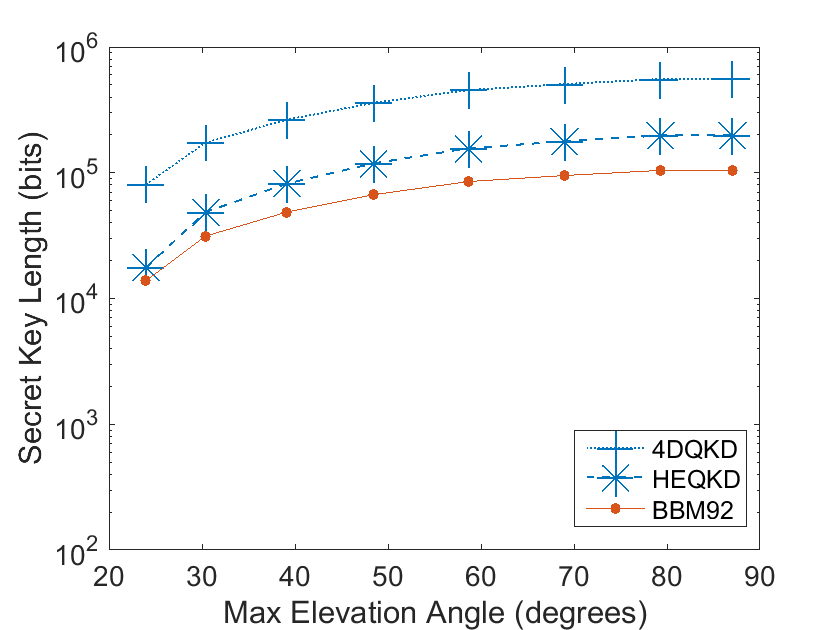}}}}\caption{LEO Protocol Comparison: (a) Optimal mean pairs per pulse $\mu$ versus the maximum elevation angle per orbital pass. When the satellite goes overhead during a pass, the maximum elevation angle is $90^{\circ}$. The channel loss (range) varied between 28 dB (400 km) and 36 dB (1000 km) in this simulation. Future system parameters used in simulations: 400-MHz laser repetition rate, 180-s orbital pass time, $10^{-6}$ background noise, and Alice's total transmission = 0.3, including detection efficiency. See Table \ref{sysparam} and see Secret-Key-Rate Simulations, Sec. \ref{SKRS}, for more details. (b) Optimal 4DQKD basis probability $p$, optimal HEQKD basis probability $p$, and optimal BBM92 parameter estimation ratio $r$, versus the maximum elevation angle per orbital pass. (c) Simulated secret key length per orbital pass for 4DQKD, HEQKD, and BBM92, versus the maximum elevation angle per orbital pass.}\label{HEQKDvsBBM92}
\end{figure}

Compared to a GEO link, the channel losses from a transmitter in LEO would be much less; however, the loss depends on the range, the separation between sender and receiver, which varies as the spacecraft flies overhead (except for a GEO orbit, where by definition the satellite is always overhead). Specifically, the instantaneous elevation angle of a LEO satellite with respect to some terrestrial observatory changes as it passes overhead, with a maximum elevation angle that varies from pass to pass ($90^{\circ}$ max elevation angle corresponding to a pass directly overhead). Fig. \ref{HEQKDvsBBM92} shows the same calculations as Fig. \ref{fksim} but now as a function of the elevation angle for a LEO satellite. In Fig. \ref{HEQKDvsBBM92}c we show the predicted secret key length per pass for 4DQKD, HEQKD, and BBM92, versus maximum elevation angle, assuming the minimum acceptable elevation angle during a pass is $20^{\circ}$ (below this we assume a reliable link cannot be established, as the effects of turbulence and scattering are much worse near the horizon). This was calculated using the optimized $\mu$, $p$, and $r$, as shown in Fig. \ref{HEQKDvsBBM92}a-b, when data from the whole pass was analyzed all at once, but not combining any data from multiple passes.

For these calculations, we used simulated orbit data for all orbital parameters, assuming a 400-km altitude and $51^{\circ}$ inclination; the range was calculated from LEO to a ground station located at $39^{\circ}$ N latitude. We used the Friis equation to estimate channel transmission, assuming a transmitting telescope diameter $D_T=0.1$ m, receiving telescope diameter $D_R=1$ m, and wavelength $\lambda=1550$ nm, with the added assumptions of a 6-dB loss for receiver telescope transmission~\cite{biswas2014optical} and  adaptive optics single-mode fiber collection efficiency~\cite{chen2015experimental}, negligible atmospheric loss, negligible loss from PAT errors, and the other the system parameters listed in Table \ref{sysparam}. We discretely integrated over the whole pass in 10-s increments; data acquisition was assumed to begin and end at 20$^{\circ}$ elevation angle, corresponding to a maximum range of about 1000 km with a minimum range of 400 km. For comparison, at those same distances, a fiber-optic cable (assuming 0.2 dB/km of attenuation) would have 80-200 dB of loss, much more than the predicted 25-30 dB loss in a free-space implementation, leading to transmission rates lower by 17 orders of magnitude! With these assumptions, a future implementation of this system in LEO could generate a secret key of substantial length in a single pass. Furthermore, the secret key length generated by HEQKD exceeds that of BBM92 by about a factor of 2 for all maximum elevation angles between $30^{\circ}$ and $90^{\circ}$.

Fig. \ref{HEQKDvsBBM92}c also shows that HEQKD is a type of hybrid between BBM92 and 4DQKD, the former always producing a secure key rate which is bounded by BBM92 and 4DQKD in this scenario. At a $90^{\circ}$ maximum elevation angle, BBM92 produces about 100 kbits/pass of secure key; in comparison, HEQKD and 4DQKD produce 200 kbits/pass and 550kbits/pass, respectively. HEQKD is able to take advantage of some of the higher dimensional benefits of 4DQKD without the increased complexity of truly implementing 4DQKD.  Although HEQKD provides a marked improvement over BBM92, the increased technical complexity of HEQKD over BBM92  would need to be factored into whether it is worth implementing HEQKD in a satellite demonstration, where size, weight, and power (SWAP) constraints are crucial. Notably, integrated optics could potentially provide a substantially lower SWAP version of a BBM92 and/or HEQKD system~\cite{Corrielli2014-tt,Xiong:15,Bassithesis}, making the SWAP difference between them negligible. For example, ultrafast low-loss thin-film lithium niobate integrated optic switches~\footnote{Note that such fast switches can permit time-bin separations less than the detector timing resolution to be resolved: the different time bins are simply switched to different detectors.} have recently achieved switching  times lower than 10 ps~\cite{Zhu:21}, which would enable time-bin separations smaller than 100 ps, achievable on a photonic chip~\cite{Zhou_2020}. Note that reducing the time-bin separation has the collateral benefit of reducing the Doppler effect as well. In this case, it is clear that the higher secret key rates provided by HEQKD are preferable.

\section{Qubit-Encoding System Engineering Considerations} \label{sec:QEsyseng}
All photon-based quantum communication implementations must decide what degree(s) of freedom will be used to encode the quantum information; these decisions will be informed by the link characteristics, e.g., time-bin encoding may be better preserved in optical fibers than polarization encoding. Such considerations are potentially even more critical when evaluating a space-earth link.

We chose to implement a hyperentangled system (as opposed to one using higher dimensional encoding in a single degree of freedom, e.g., only time bins~\cite{islam2017provably} or orbital angular momentum (OAM)~\cite{mirhosseini2015high, bouchard2018experimental,PhysRevApplied.11.024070,d2012complete,PhysRevLett.113.060503}) to investigate the feasibility and advantages of multiple photonic degrees of freedom in a single demonstration. Polarization was chosen for its relative ease of manipulation, and both polarization and time bins for their relative robustness in the atmospheric channel. In contrast, OAM modes are corrupted by turbulence, limiting their applicability in this scenario~\cite{PhysRevApplied.14.064031}; furthermore, the whole mode must be collected, not just part of it, or the OAM states cannot be reliably distinguished. In a satellite-based implementation, time bins are susceptible to the Doppler shift from the satellite's relative motion, which affects not only the phase between the time bins but also their separation. However, we have shown that it is straightforward to correct for this in some applications, e.g., QKD (see Sec. \ref{sec:Dopp}) and superdense teleportation~\cite{chapman2019time}. Finally, we note that frequency entanglement and energy-time entanglement with a continuous-wave pump are other possibilities for further research~\cite{imany201850}.

To successfully execute a quantum communication protocol, like quantum key distribution, in a space-to-Earth channel and for the conclusions from our lab-based demonstration to be valid, there are several pertinent requirements that need to be satisfied. Some are unique and specific to polarization encoding, others to time-bin encoding, and still others due to a satellite-to-ground link. Here we will sketch the relevant requirements; in Appendix~\ref{capreq} they are shown in more detail. For polarization encoding, the sender and receiver need shared polarization bases. This is often accomplished by active polarization compensation (to correct for rotations and phase shifts produced by the sending and receiving optics). Polarization encoding obviously also requires high-extinction polarizers and/or polarizing beamsplitters. Time-bin encoding requires an adaptive optics system at the ground receiver station (to correct for wave-front distortions by the atmosphere and allow for single-mode fiber coupling of the received light)~\footnote{As an alternative, one can engineer a multi-mode robust unbalanced interferometer; this has been previously demonstrated several ways, achieving visibilities in excess of 98\%~\cite{PhysRevApplied.13.024047,PhysRevA.97.043847}. Unfortunately, the accepted sky background will be much higher in this case.}, and active phase stabilization (to phase stabilize the time bins from the Doppler shift, see Sec. \ref{sec:Dopp}, and from other variations, e.g., satellite vibrations or laboratory temperature fluctuations). Note that all preceding requirements were satisfied in some way, sometimes trivially, in our lab, so they are not additional requirements to our demonstration, but necessary requirements for a successful space-to-ground demonstration that would validate the conclusions of our lab-based study. More general considerations for a satellite-to-ground demonstration include sky-background-light filtering, clock distribution for coincidence post-processing, space-qualified single-photon detection, space-qualified entangled-photon source and pump laser. These requirements and the others listed in Appendix \ref{capreq}, need to be fulfilled for the conclusions of this paper to be valid; fortunately, they can already be fulfilled with today's technology, implying the feasibility of successfully implementing HEQKD, BBM92, and related quantum protocols.

QKD requires only measurement and detection of received photons. In contrast, other quantum communication protocols like teleportation~\cite{quanttel} and entanglement swapping~\cite{entswappaper} may require transmitted photons to interfere with other photons, e.g., for Bell-state analysis~\cite{PhysRevA.51.R1727}. The need for these interfering photons to be intrinsically indistinguishable from each other leads to other requirements and considerations. Importantly, the photons pairs produced via SPDC must be spectrally unentangled from one another so their individual states in all other degrees of freedom are separable and pure~\cite{quanttel}. The Doppler shift from the satellite's motion (see Sec. \ref{sec:Dopp}) presents several new challenging considerations, as it affects all frequencies, including the repetition rate (if there is one) and the optical spectra center. This Doppler shift will therefore reduce the temporal and spectral overlap of the photons entering the Bell-state analyzer, assuming the photons are emitted by platforms in relative motion. Since all spectral content of the photons is affected, engineering solutions need to simultaneously consider the Doppler shift's effect on all frequencies and not just on one frequency band, e.g., the repetition rate. For example, the phase stabilization we introduce in Sec.~\ref{sec:Dopp} corrects for the relative separation of the time bins; because the path-length difference in the analysis interferometer is changed to match the separation of the time bins. In this case, there is no change in the measured relative phase, even though the frequency of the photons is also shifted. However, if instead one is trying to interfere this frequency-shifted photon with one produced by an identical source on a non-relatively moving platform (or trying to couple it into a quantum memory whose absorption transition frequency matches the photon from the source when it is stationary and the linewidth is similar or smaller than the Doppler shift), the different frequencies will reduce the interference (or the coupling probability) unless one can make them indistinguishable again, e.g., by applying a continuously tunable frequency shift on the photon (or the quantum memory) or using filtering, and employ phase stabilization for the time-bin qubit. The $\pm4$-GHz shift experienced by our 200-GHz-bandwidth SPDC photons when emitted by a LEO satellite moving toward/away from the receiver is negligible, but photons designed to couple to a typical quantum memory, with bandwidth 0.01-1 GHz, would definitely require such Doppler compensation in order to enable, e.g., coupling to a quantum memory or entanglement swapping.
\section{Discussion}
We have presented a direct comparison between HEQKD and BBM92. Using finite-key analysis and the optimizations in Fig. \ref{fksim}-\ref{HEQKDvsBBM92}, we found HEQKD allows for secret key generation at higher rates than does BBM92, including in geostationary orbit, except at the highest losses due to more detectors collecting more background and dark count detection events. Moreover, HEQKD allows for secure key generation with higher key rates than BBM92 when the losses are less extreme, e.g., in low-Earth orbit or medium-Earth orbit, at all relevant maximum elevation angles. This comparison was done with all system parameters being equal except choosing the optimal $\mu$ and $p$ for HEQKD and the optimal $\mu$ and $r$ for BBM92. We did not optimize the secure key rate by changing the basis choice probability for BBM92 because our system uses a non-polarizing beamsplitter to set the ratio to 50\% (which is important for our other protocol demonstrations with this system \cite{chapman2019time}). Our security analysis actually used data from both bases for key generation and error checking and we did optimize the parameter estimation ratio for maximum secret key length, which is usually why one would optimize the basis choice probability in other analysis frameworks where one basis is used for key generation, and another is used for error checking. Furthermore, the error rates of the two bases in BBM92 are very similar (1\% and 2\%) and a percent change in QBER in this case has little effect on the secret key length.

From these measurements and analyses, we project this system (with the same QBER but enhanced repetition rate and efficiency as we used in our simulations) would be suitable for operation in the channel between space and Earth, and would be able to generate a sizable, usable secret key within a single orbital pass. Furthermore, we find that the use of time-bin qubits in general should be feasible for a channel that includes an orbiting platform, assuming one compensates for the adverse effect from the Doppler shift, as we have done. The background detection rates assumed in this analysis are typically found during nighttime operation with spatial and tight spectral filtering, but with improved adaptive optics daytime operation without tight spectral filtering is possible~\cite{PhysRevApplied.16.014067}. This would significantly increase the number of eligible passes for secret key generation. Additionally, after the aforementioned compensation systems have been activated, but prior to protocol execution, a phase calibration step is necessary for bases which include superpositions of time bins and/or polarizations, so that Alice and Bob are indeed measuring in the same bases. Implementing such systems with high precision is readily achievable with current technology and should enable QBER levels comparable to or maybe even better than the values measured in our laboratory. Therefore, we conclude that useful quantum key distribution from a satellite, in LEO, MEO, or even GEO, should be feasible with existing technology. Moreover, modulo the considerations discussed in Sec.~\ref{sec:QEsyseng}, the polarization and time-bin encoded photons could also be used in other protocols such as teleportation and entanglement swapping.

\section{Acknowledgments}
Thanks to Michael Wayne and Kristina Meier for discussions regarding design of the time-bin sorting circuit, and to Chris Chopp for assistance in prototyping and printed-circuit-board layout of the time-bin sorting circuit. Thanks to MIT-Lincoln Laboratory for the orbital simulation calculations. J.C.C. and P.G.K. acknowledge support from NASA Grant No. NNX13AP35A and NASA Grant No. NNX16AM26G. J.C.C. acknowledges support from a DoD, Office of Naval Research, National Defense Science and Engineering Graduate Fellowship (NDSEG) and from the DOE Office of Cybersecurity Energy Security and Emergency Response (CESER) through the Cybersecurity for Energy Delivery Systems (CEDS) program. C.C.W.L. acknowledges support from NUS startup grant R-263-000-C78-133/731 and CQT Fellow Grant. This work was partially performed at Oak Ridge National Laboratory (ORNL). ORNL is managed by UT-Battelle, LLC, under Contract No. DE-AC05-00OR22725 for the DOE. All authors contributed to experiment design and wrote manuscript.  J.C.C. carried out all experiments and data analysis, including upgrading the optical and detection system, and constructing the time-bin sorting circuit.  C.W.L. carried out theoretical security analysis, and C.W.L and J.C.C. wrote secret-key simulations.
%\section{Author Contributions}

\appendix
\section{HEQKD and BBM92 Finite Key Analysis}\label{FKA}

\subsection{Preliminaries: Models} 
In the following, we will analyze the finite-key security of BBM92 (also called entanglement-based BB84) and our high-dimensional QKD protocol using two sets of two mutually unbiased bases. The security proof technique relies on the entropic uncertainty relation~\cite{tightanalysis}. In both of the above protocols, we assume that entanglement is generated in Alice's laboratory using a non-deterministic pair source, e.g., a spontaneous parametric down-conversion (SPDC) or four-wave mixing source, where a non-linear crystal is used to split a strong laser beam into pairs of correlated photons~\cite{Ma2007}. Quantum mechanically, we can write the output state (using vector representation) of such a qubit-entangled source as
\be
\ket{\Psi}_{AB}\equiv\cosh^{-2}(\chi)\sum_{n=0}^\infty \sqrt{n+1} \tanh^n(\chi)\ket{\psi_n}_{AB},
\ee
where
\be
\ket{\psi_n}_{AB}\equiv \frac{1}{n+1}\sum_{k=0}^{n} (-1)^k\ket{n-k,k}_A\otimes\ket{n-k,k}_B.
\ee
We also assume that the photons pairs emitted are indistinguishable from one another, which can be satisfied when the pump pulse width is smaller than the downconversion coherence time~\cite{takesue2010effects}. The probability to get $n$ photon-pairs in a given pump pulse is
\be
P_n\equiv \frac{(n+1)\gamma^n}{(1+\gamma)^{n+2}}, 
\ee
where $\gamma\equiv\sinh^2(\chi)$ is related to the pump power of the laser; $2\gamma=\mu$ is the mean number of pairs per pump pulse.

To model the detection rates (which are needed for the simulation), we define $\eta_A$ and $\eta_B$ to be the overall detection efficiencies for Alice and Bob, respectively. Note that $\eta_A$ and $\eta_B$ include all the losses due to the quantum channel, coupling loss, and detection inefficiency. Using this definition, it is easy to see that the probability of observing a coincident detection (i.e., at least one click on each side) given the emission of an $n$ photon-pair state, $\ket{\psi_n}$, is
\be
\eta_n=[1-(1-\eta_A)^n][1-(1-\eta_B)^n].
\ee
In addition, we define the \emph{yield} of an $n-$photon pair to be $Y_n$, which is the conditional probability that a coincident detection is observed if the source emits $n-$photon pairs:
\be
Y_n=[1-(1-\xi_A)(1-\eta_A)^n][1-(1-\xi_B)(1-\eta_B)^n],
\ee
where $\xi_A$ ($\xi_B$) is the probability of observing a noise or background count on Alice's (Bob's) side per pump pulse~\cite{Ma2007}. For example, in the case of zero photon emission, we have $Y_0=\xi_A\xi_B$.
Using the above models, the overall coincidence detection probability per pump pulse using non-photon-number-resolving detectors is
\be
R=\sum_{n=0}^\infty P_nY_n = 1- \frac{1-\xi_A}{(1+\eta_A\gamma)^2}-\frac{1-\xi_B}{(1+\eta_B\gamma)^2}+\frac{(1-\xi_A)(1-\xi_B)}{(1+\eta_A\gamma + \eta_B\gamma-\eta_A\eta_B\gamma)^2}.
\ee

To estimate the probability of detecting an error per pump pulse, we look at the effects of measurement system errors, background light, and photon loss, including some effects from multiple-pair detection events (which require the assignment of a random bit~\cite{lutkenhaus1999quantum}). We needed to derive this estimate to have a rate formula that was suitable for $d=4$, since previous work on this topic~\cite{Ma2007} was not easily generalized to higher dimensions. Here we derive a $d$-dimensional-system estimate which gives an upper bound on the error rate, at least for the $d=2$ case when compared to the formula for $E_{\lambda}Q_{\lambda}$, derived in~\cite{Ma2007} (which is an exact formula for the expected error rate for a system of that dimension).

We estimate the error from background events for 3 different cases: Alice (Bob) loses all her (his) source photons but detects a background event and Bob (Alice) detects at least one source photon (assuming there is no error); or both Alice and Bob lose all their source photons and they both detect a background event. Conditional on any of these events happening, there is an error probability of $e_0=(d-1)/d$ since the background photons are assumed to occur randomly at each detector. Summing the errors from these three cases, we estimate the total error probability for background-related errors to be
\begin{align}
E_{b}=e_0\sum_{n=0}^{\infty}P_n&\Bigg\{\sum_{k=1}^{n}\Big[\binom{n}{k}\eta_A^k(1-\eta_A)^{n-k}\xi_B(1-\eta_B)^n\Big]\nonumber\\
&+\sum_{l=1}^{n}\Big[\binom{n}{l}\eta_B^l(1-\eta_B)^{n-l}\xi_A(1-\eta_A)^n\Big]\nonumber\\
&+\xi_A\xi_B(1-\eta_A)^n(1-\eta_B)^n\Bigg\}.\label{ebgn}
\end{align}
The first term/top line (second term/middle line) in the curly braces corresponds to the probability that Alice (Bob) detects at least one source photon and Bob (Alice) detects only a background photon; the last term/bottom line is the probability that Alice and Bob each only detect a background photon.

In addition to these background-related errors, there are some situations where all source photons produced are detected, e.g., when the source produces one pair, and we detect one pair. Errors in those cases arise from imperfections in the measurement system. Also, for $n>1$, errors arise from the uncorrelated nature of certain terms in the higher-order photon-pair wavefunction Eqn. \ref{psiN}, from~\cite{kok2000postselected}. These errors result in multiple source photons being detected simultaneously on different detectors of the same side; contrary to the initial impulse to simply discard these cases, here we assign a random bit to the key (to avoid the possibility of `bit forcing' by an eavesdropper)~\cite{lutkenhaus1999quantum}.

The terms in $\ket{\Psi^n}$, where all photons created on a given side are in the same mode (for $d=2$, assuming type-I entangled pair generation, these terms are $\ket{2,0}_A\otimes\ket{2,0}_B$ and $\ket{0,2}_A\otimes\ket{0,2}_B$), occur with net probability $dN_n^2$, where $N_n$ is the normalization factor of $\ket{\Psi^n}$ (for $d=2$, $N_n^2=1/(n!(n+1)!)$ ~\cite{kok2000postselected}). To calculate $N_n$ for HEQKD (d=4), we need to generalize~\cite{kok2000postselected} to the case of four measurement outcomes instead of two. In this case, the Hamiltonian is 

\be
H=i\kappa(a_0^{\dagger}b_0^{\dagger}+a_1^{\dagger}b_1^{\dagger}+a_2^{\dagger}b_2^{\dagger}+a_3^{\dagger}b_3^{\dagger})+ \text{H.c.} \text{,}
\ee
where $\kappa$ is the product of the pump amplitude and the coupling constant between the electromagnetic field and the crystal, and H.c. means Hermitian conjugate. The operators $a_i^{\dagger}$, $b_i^{\dagger}$ and $a_i$, $b_i$ are creation and annihilation operators for the 4 basis states in our measurement system, e.g., $a_0^{\dagger}\ket{0}=\ket{Ht_1}$, $a_1^{\dagger}\ket{0}=\ket{Vt_2}$, $a_2^{\dagger}\ket{0}=\ket{Vt_1}$, and $a_3^{\dagger}\ket{0}=\ket{Ht_2}$. Using the multinomial theorem, we find the state of $n$ entangled photon pairs to be
\be
\ket{\Psi^n}=N_nL_+^n\ket{0}=N_n\sum_{k_3=0}^{n}\sum_{k_2=0}^{n-k_3}\sum_{k_1=0}^{n-k_3-k_2}\Bigg[n!\ket{k_0,k_1,k_2,k_3;k_0,k_1,k_2,k_3}_{a_0,a_1,a_2,a_3,b_0,b_1,b_2,b_3}\Bigg],\label{psiN}
\ee
where we define $L_+\equiv a_0^{\dagger}b_0^{\dagger}+a_1^{\dagger}b_1^{\dagger}+a_2^{\dagger}b_2^{\dagger}+a_3^{\dagger}b_3^{\dagger}$ and $k_0\equiv n-k_3-k_2-k_1$. $N_n$ is a normalization factor so that $\bracket{\Psi^n}{\Psi^n}=1$, where
\be
\frac{1}{N_n^2}=(n!)^2 \binom{n+4-1}{4-1}=\frac{n!(n+3)!}{6}
\ee
is $(n!)^2$ multiplied by the composition of $n$ into $4$ parts ($k_0$, $k_1$, $k_2$, and $k_3$) where zero is allowed~\cite{MWcomp}. The terms in Eqn. \ref{psiN} where all $n$ photons produced on a given side (i.e., $a$ or $b$) are in the same mode (i.e., 0, 1, 2, or 3) only happen in our system when all $n$ photon pairs produced are in the same mode (e.g., $\ket{n,0,0,0;n,0,0,0}_{a_0,a_1,a_2,a_3,b_0,b_1,b_2,b_3}$). Thus, for events where all produced photons are detected and all $n$ photons produced on a given side are in the same mode, some undetected measurement system errors happen when all the photons on one side are incorrectly sorted to the \emph{same} wrong detector, which happens with probability $(e_d/(d-1))^nd(n!)^2N_n^2$, where we assume that errors are evenly distributed among all the states in a basis and $e_d$ is the intrinsic error probability in that basis, conditional on a coincidence detection. The rest of the errors when all created photons are detected occur when they are detected by at least two detectors (thereby giving a clear indicator that there were multiple pairs created); these incur an error probability of $e_0$ since we again need to assign a random bit value. Therefore, the total error probability when all produced photons are detected is
\be
E_{\ket{\Phi^n}}=\sum_{n=1}^{\infty}P_n\eta_A^n\eta_B^n\Big(\frac{e_d}{d-1}\Big)^nd(n!)^2N_n^2+\sum_{n=2}^{\infty}P_n\eta_A^n\eta_B^n\Big(1-\Big(\frac{e_d}{d-1}\Big)^nd(n!)^2N_n^2\Big)e_0.
\ee

Now we consider the more likely situation in which some of the created photons are \emph{not} detected; we calculate the error probability when at least two photon pairs are created and at least one photon on each side is detected (possibly not from the same pair) and we assume that all events detected are either uncorrelated or have multiple detections; in either case we must assign a random bit. The error probability for these events is

\be
E_{\text{MPE}}=e_0\sum_{n=2}^{\infty}P_n\Bigg[\sum_{i=1}^{n}\Big[\binom{n}{i}\eta_A^i(1-\eta_A)^{n-i}\Big]\sum_{j=1}^{n}\Big[\binom{n}{j}\eta_A^j(1-\eta_A)^{n-j}\Big]-\eta_A^n\eta_B^n\Bigg].
\ee

Here we slightly overestimate the error probability by including, as errors, the cases where all photons detected happened to go to the same detector but \emph{not} cause an error. Thus, for a $d$-dimensional system, our estimated total error probability per pump pulse is
\be
Q_{d,\tn{est}}\leq\frac{E_{b}+E_{\ket{\Phi^n}}+E_{\text{MPE}}}{R}.
\ee
Since the probability of higher order $n$-photon pair events diminishes rapidly as $n$ increases, we think this over-counting produces a relatively tight upper bound on the true $d=4$ multi-photon error probability.

\subsection{Security Bound for BBM92}

Having defined the quantum channel and source models, we can derive a bound on the extractable key length (denoted by $\ell$) for a given post-processing block size (the number of raw bits we collect in one execution of the protocol). 

To further model the security of the QKD protocol, we consider a ($d=2$) QKD protocol with two mutually unbiased bases. Alice's and Bob's bases are written as $\{\mathsf{A}_i\}_{i=1}^2$ and $\{\mathsf{B}_i\}_{i=1}^2$, respectively. Assuming Alice and Bob are each operating in a two-dimensional Hilbert space with computational basis given by $\mathsf{Z}=\{\ket{i}\}_{i=1}^2$, their measurement bases are defined as $\mathsf{A}_1\equiv\mathsf{Z}$, $\mathsf{A}_2\equiv\{(\ket{0}\pm\ket{1})/\sqrt{2} \}$, and similarly for Bob. The bases are uniformly chosen (with 1/2 probability each) and the raw key is generated from both bases. That is, the raw key is randomly sampled from the measurement data (of size $N$); thus, the size of the raw key is fixed to some positive integer $m=n+k$, where $k=m(1-r)$ is the amount of raw key used for parameter estimation, $n=mr$ is the amount of raw key left for key generation, and $r$ is the security parameter estimation ratio. Following standard security definitions~\cite{composability}, we say that the QKD protocol is $\eps$-secure if it is both $\eps_\tn{sec}$-secret and $\eps_\tn{cor}$-correct. The protocol is called $\eps_\tn{sec}$-secret if the joint state of the output secret key (say on Alice's side) and the adversary's total quantum information is statistically indistinguishable from the ideal output state except with some small probability $\eps_\tn{sec}$. The ideal output state is an output key which is uniformly random (in the key space) and completely independent of the adversary's total information. For the second condition, the protocol is called $\eps_\tn{cor}$-correct if the output secret keys on Alice and Bob's sides are identical except with some small probability $\eps_\tn{cor}$.

The starting point of our security analysis is to ask how many secret bits can be extracted from Alice's raw key ${X}$ (of size $m$) given ${E}$ (Eve's total information about the QKD system). To this end, we use the quantum leftover-hash lemma~\cite{tomhayinfo} to bound the secret key length, $\ell$, giving
\be \label{eq:length}
\ell= \max_{\beta\in(0,\eps_\tn{sec}/2]}\left\lfloor H_\tn{min}^{\eps_\tn{sec}/2-\beta}(X|E)+4\log_2\beta-2 \right\rfloor,
\ee 
where $H_\tn{min}^{\eps_\tn{sec}/2-\beta}$ is the smooth min-entropy of $X$ given $E$ (see Ref.~\cite{minentropy} for more details) and $\beta$ is a constrained optimization parameter.

Using the fact that there exists a squashing model (a theoretical argument that maps higher photon number states to a qubit state)~\cite{Squashing2008} for two mutually unbiased measurements (it applies regardless of whether the implementation uses active or passive basis choice), we can bound the min-entropy using the entropic uncertainty relation~\cite{tightanalysis}, assuming the measurements on Alice's side are mutually unbiased (e.g., no polarization misalignment at the measurement level). More specifically, we have
\be
H_\tn{min}^{\eps_\tn{sec}/2-\beta}(X|E) \geq n(1-h_2(Q_{2,\tn{est}}+\Delta(n,k,\beta))) -\tn{Leak}_{\tn{EC}}^{2D}-\log_2\frac{2}{\eps_\tn{cor}}\text{ and}
\ee
\be
\tn{Leak}_{\tn{EC}}^{2D}= 1.12nh_2(Q_{2,\tn{est}} )\text{,}
\ee
where 
\be
\Delta(n,k,\beta)\equiv\sqrt{\frac{n+k}{nk}\frac{k+1}{k}\ln\frac{1}{\beta}},
\ee
\be
h_2(x)\equiv-x\log_2{(x)}-(1-x)\log_2{(1-x)}
\label{binaryentropy}
\ee
is the binary entropy, $\Delta$ is the statistical noise due to finite statistics of the amount of raw key used for parameter estimation $k$ and the raw key left for key generation $n$, and $\tn{Leak}_{\tn{EC}}^{2D}$ and $\log_2 {2}/{\eps_\tn{cor}}$ are the information leakages due to error correction and verification, respectively. 
Putting everything together, we get 
\be
\ell^{2D} = \max_{\beta \in (0,\eps_\tn{sec}/4)} \left\lfloor n_\tn{ext}^{2D} +4\log_2 \beta -2 \right\rfloor, 
\ee
where
\begin{equation}
n_\tn{ext}^{2D}\equiv n(1-h_2(Q_{2,\tn{est}}+\Delta(n,k,\beta))) -\tn{Leak}_{\tn{EC}}^{2D}-\log_2\frac{2}{\eps_\tn{cor}}.
\end{equation}

$n_\tn{ext}^{2D}$ means the maximum theoretical length one can extract with the protocol, while $\ell^{2D}\leq n_\tn{ext}^{2D}$  is the final output of the QKD system in practice.
\subsection{Security Bound for HEQKD}
We consider an entanglement-based ($d=4$) QKD protocol with four measurement bases.
Alice's and Bob's bases are written as $\{\mathsf{A}_i\}_{i=1}^4$ and $\{\mathsf{B}_i\}_{i=1}^4$, respectively, and we suppose that their reference frames and measurements are aligned, i.e., $\mathsf{A}_i=\mathsf{B}_i$ for $i=1,2,3,4$. Assuming Alice and Bob are each operating in a four-dimensional Hilbert space with computational basis given by $\mathsf{Z}=\{\ket{i}\}_{i=1}^4$, their measurement bases are defined as $\mathsf{A}_1\equiv\mathsf{Z}$, $\mathsf{A}_2\equiv\{(\ket{0}\pm\ket{1})/\sqrt{2},(\ket{2}\pm\ket{3})/\sqrt{2} \}$, $\mathsf{A}_3\equiv\{(\ket{0}\pm\ket{2})/\sqrt{2},(\ket{1}\pm\ket{3})/\sqrt{2} \}$, and $\mathsf{A}_4\equiv\{(\ket{0}+\ket{1}+\ket{2}-\ket{3})/2$, $(\ket{0}+\ket{1}-\ket{2}+\ket{3})/2$, $ (\ket{0}-\ket{1}+\ket{2}+\ket{3})/2$, $ (\ket{0}-\ket{1}-\ket{2}-\ket{3})/2 \}$, and similarly for Bob. It can be easily checked that bases 1 and 4 and bases 2 and 3 are mutually unbiased. Bases 1 and 2 are each chosen with probability $p$, while bases 3 and 4 are each chosen with probability $q$. Hence, we have that $2p+2q=1$, or $q=1/2-p$.
After the measurement phase, Alice and Bob perform sifting (via public communication) to identify successful events according to their basis choices. We denote these sets by $\mathcal{S}_{i,i'}$ and their respective lengths by $m_{i,i'}$. For example, the data belonging to set $\mathcal{S}_{1,3}$ are the events in which Alice and Bob chose basis 1 and 3, respectively, and each detected one photon (though the results are not necessarily correct). 

In our analysis, we focus on processing Alice's measurement data to extract her secret key; our analysis can easily be reversed for Bob's measurement data. We partition her data into four sets, namely, sets containing events in which Alice chooses either basis 1 or basis 2 (which data comprises the raw key) and sets containing events in which Alice chooses either basis 3 or basis 4 (partially used to determine Alice's QBER). Alice's data in basis 1 will be paired with Bob's data from basis 1, 2 and 3, and data in basis 2 will be paired with Bob's data from basis 1, 2 and 4. Note that common bases should ideally generate perfectly correlated data (2 bits per measurement) while bases that are not mutually unbiased and not common (e.g., basis 1 and basis 2) should ideally generate partially correlated data (1 bit per measurement). $\mathcal{S}_{3,3}$ and $\mathcal{S}_{4,4}$ are used for error estimation; $\mathcal{S}_{3,4}$ and $\mathcal{S}_{4,3}$ are unusable because it is insecure to extract secure key from bases used for error checking, in this protocol. There is no correlation when Alice and Bob choose mutually unbiased bases, e.g., $\mathcal{S}_{1,4}$ or $\mathcal{S}_{2,3}$, so those events are sifted out of the raw key in the reconciliation phase of the public discussion between Alice and Bob. See Table \ref{HEQKDbasistab} for the complete list of pairings.
\begin{table}
\caption{HEQKD Bases: All basis combinations in the HEQKD protocol and their effect on key generation. MUB = mutually unbiased bases; Bits/Photon, in this case, means bits of raw key per sifted coincident photon pair.}
%\centerline{\includegraphics[scale=0.9]{basiskeytablev7.png}}
\label{HEQKDbasistab}
\begin{center}
	\begin{tabular}{ c c c c}
	\hline
	\hline
		  Alice Basis & Bob Basis & Bits/Photon & Use \\ \hline

		1&1&2&Data\\
		1&2&1&Data\\
		1&3&1&Data\\
		1&4&-&MUB\\
		\hline
		2&1&1&Data\\
		2&2&2&Data\\
		2&3&-&MUB\\
		2&4&1&Data\\
		\hline
		3&1&1&Data\\
		3&2&-&MUB\\
		3&3&2&Error Estimation\\
		3&4&1&Unusable\\
		\hline
		4&1&-&MUB\\
		4&2&1&Data\\
		4&3&1&Unusable\\
		4&4&2&Error Estimation\\
	\hline
	\hline		
			\end{tabular}
\end{center}
\end{table}
To compute the finite-key security of Alice's data, we first need to introduce some random variables to capture the random behavior of the measurements. To that end, let $\rv{X}_1$ be the random string of length $n_1=m_{1,1}+m_{1,2}+m_{1,3}$ describing Alice's measurement outcomes when she chooses basis 1. Likewise, for the case when Alice chooses basis 2 we write $\rv{X}_2$ to denote the random string of $n_2=m_{2,1}+m_{2,2}+m_{2,4}$. Recall, the lengths of the sifted events when Alice measures in basis $i$ and Bob measures in basis $i'$ are denoted by $m_{i,i'}$. Our immediate goal now is to show that it is possible to extract a secret key of length $\ell>0$ from $\rv{X}_1\rv{X}_2$ if certain experimental conditions are met.

The starting point of our security analysis is to ask how many secret bits can be extracted from Alice's raw key $\rv{X}$ (of size $n$) given $\rv{E}^+$ (Eve's total information about the overall joint state shared between Alice and Bob, including the classical communication sent by Alice to Bob). To this end, we use the quantum leftover-hash lemma~\cite{tomhayinfo} to determine the secret key length, $\ell$, giving
\be \label{eq:lengthd4}
\ell= \max_{\beta\in(0,\eps_\tn{sec}/2]}\left\lfloor H_\tn{min}^{\eps_\tn{sec}/2-\beta}(\rv{X}_1\rv{X}_2|\rv{E}^+)+4\log_2\beta-2 \right\rfloor,
\ee 
where the left-hand term in the floor function is the smooth min-entropy of $\rv{X}_1\rv{X}_2$ given $\rv{E}^+$ (see Ref.~\cite{minentropy} for more details). We can further break up the min-entropy term into two parts by using a chain-rule inequality for smooth min-entropies,
\be
H_\tn{min}^{\eps_\tn{sec}/2-\beta}(\rv{X}_1\rv{X}_2|\rv{E}^+) \geq H_\tn{min}^{\bar{\eps}}(\rv{X}_1|\rv{X}_2\rv{E}^+) + H_\tn{min}^{\bar{\eps}}(\rv{X}_2|\rv{E}^+)+\log\left(1-(1-\bar{\eps}^2)^{1/2}\right),
\ee
where $\bar{\eps}\equiv\eps_\tn{sec}/6-\beta/3$. To further simplify the analysis, we assume that $\rv{X}_1$ and $\rv{X}_2$ are independent. In the experiment, this assumption can be achieved by having Alice use a pulsed pump laser (with a pulse spacing larger than the coherence time of the photon pairs) to prepare highly entangled photon pairs (which are independent from each other between different laser pulses), measure one half of each entangled photon pair, and send the other half to Bob via the quantum channel. In this case, each pump pulse generates pair(s) independent from the pair(s) created during other channel uses, i.e., other pump-laser pulses. This procedure in effect produces random outcomes in each run, which implies $\rv{X}_1$ and $\rv{X}_2$ are independent variables. With this assumption, we have  
\be
H_\tn{min}^{\eps_\tn{sec}/2-\beta}(\rv{X}_1\rv{X}_2|\rv{E}^+) \geq H_\tn{min}^{\bar{\eps}}(\rv{X}_1|\rv{E}^+) + H_\tn{min}^{\bar{\eps}}(\rv{X}_2|\rv{E}^+)+\log\left(1-(1-\bar{\eps}^2)^{1/2}\right),
\ee
where now the smooth entropy terms $H_\tn{min}^{\bar{\eps}}(\rv{X}_1|\rv{E}^+)$ and $H_\tn{min}^{\bar{\eps}}(\rv{X}_2|\rv{E}^+)$ can be treated independently. To translate these terms into expressions that can be bounded using experimental data, we use a version of entropic uncertainty relations for two $d=4$ mutually unbiased bases to get
\begin{eqnarray}
H_\tn{min}^{\bar{\eps}}(\rv{X}_1|\rv{E}^+) &\geq& 2n_1 - H_\tn{max}^{\bar{\eps}}(\rv{T}_1|\rv{T}'_1),\\
H_\tn{min}^{\bar{\eps}}(\rv{X}_2|\rv{E}^+) &\geq& 2n_2 - H_\tn{max}^{\bar{\eps}}(\rv{T}_2|\rv{T}'_2),
\end{eqnarray}
where $\rv{T}_1$ and $\rv{T}'_1$ are Alice's and Bob's measurement outcomes corresponding to basis 4, and $\rv{T}_2$ and $\rv{T}'_2$ are the measurement outcomes corresponding to basis 3. Here, we assume that the measurements are acting locally on a four-dimensional Hilbert space. This assumption is valid if there is a universal squashing model~\cite{fung2011universal}, or if the entangled photon source produces independent pairs when there is a multi-pair event. Assuming Alice's and Bob's error probabilities within a given basis are uniformly distributed (i.e., they can be modeled by a depolarizing channel~\cite{wang2009entanglement}), then we have that

\begin{eqnarray}
H_\tn{min}^{\bar{\eps}}(\rv{X}_1|\rv{E}^+) &\geq& n_1(2 - h_4(Q_{4,4,4,\tn{est}}+\nu(n_1,m_{4,4},\bar{\eps}))),\\
H_\tn{min}^{\bar{\eps}}(\rv{X}_2|\rv{E}^+) &\geq& n_2(2 - h_4(Q_{3,3,4,\tn{est}}+\nu(n_2,m_{3,3},\bar{\eps}))),
\end{eqnarray}
where $Q_{i,i'}^{4D}$ is the observed error rate conditioned on Alice and Bob choosing basis $i$ and $i'$,
\be
h_4(x)\equiv-x\log_2{(x)}-(1-x)\log_2{(1-x)}+x\log_2{(3)}
\label{shannonentropy}
\ee
is the Shannon entropy, and 
\be
\nu(n,k,\eps)= \sqrt{\frac{(n+k)(k+1)\ln(2/\eps)}{nk^2}}
\ee
is the statistical error due to finite sampling, for $n$ bits used for key generation, $k$ bits used for error estimation, and $\eps$ for the $\eps$-security level. Note that in the infinite key limit this term goes to zero. 

Putting everything together, we can now establish a lower bound on $H_\tn{min}^{\eps_\tn{sec}/2-\beta}(\rv{X}_1\rv{X}_2|\rv{E}^+)\geq n_\tn{ext}$:

\begin{multline}
n_\tn{ext}^{4D}\equiv n_1(2 - h_4(Q_{4,4,4,\tn{est}}+\nu(n_1,m_{4,4},\bar{\eps})))+n_2(2 - h_4(Q_{3,3,4,\tn{est}}+\nu(n_2,m_{3,3},\bar{\eps})))-\tn{Leak}_{\tn{EC}}^{4D}-\log_2\frac{2}{\eps_\tn{cor}},
\end{multline}
\begin{align}
\tn{Leak}_{\tn{EC}}^{4D}= &1.2n_1h_4(\text{min}[0.75,p^2Q_{1,1,4,\tn{est}} + p^2Q_{1,2,4,\tn{est}} + pqQ_{1,3,4,\tn{est}} ] )+\notag\\
& 1.2n_2h_4(\text{min}[0.75,p^2Q_{2,1,4,\tn{est}} + p^2Q_{2,2,4,\tn{est}} + pqQ_{2,4,4,\tn{est}} ] )\text{,}
\end{align}
where $\tn{Leak}_{\tn{EC}}^{4D}$ and $\log_2 {2}/{\eps_\tn{cor}}$ are the leakages due to error correction and verification, respectively. Finally, we find 
\be
\ell^{4D} = \max_{\beta \in (0,\eps_\tn{sec}/4)} \left\lfloor n_\tn{ext}^{4D}+4\log_2 \beta -2 \right\rfloor. 
\ee

$n_\tn{ext}^{4D}$ means the maximum theoretical length one can extract with the protocol, while $\ell^{4D}\leq n_\tn{ext}^{4D}$  is the final output of the QKD system in practice.
\section{Capability Requirements for HEQKD and BBM92}
\label{capreq}
Here we outline various system requirements for a quantum modem in space and a ground station to implement BBM92 and HEQKD. Most requirements are generally applicable to any other polarization- and/or time-bin-encoding demonstration from space-to-Earth as well, though as discussed in Sec.~\ref{sec:QEsyseng} there may be additional requirements as well, e.g., to implement teleportation and entanglement swapping from platforms in relative motion. As such, we itemize, with \textbf{P}, \textbf{T}, or \textbf{S}, the requirements which pertain to polarization, time bins, or just space-to-Earth considerations, respectively.  HEQKD requirements that \emph{differ from BBM92} are \emph{italicized}. For the quantum modem in space, temperature stability is not explicitly called out here, but many requirements are only satisfied in a temperature-stabilized environment or with very thoughtful construction. Similarly, we implicitly assume any final system to be launched into space would be designed to minimize misalignment effects from vibrations, e.g., from launch and in orbit.
\subsection{Capability Requirements for BBM92}
 Given a system with a functioning polarization entanglement source, here are the unique system requirements to implement BBM92 in a quantum modem in space:

Alice's mutually unbiased basis measurement system in space
\begin{itemize}
\item[\textbf{P:}]  Needs high-extinction ratio polarization beamsplitters ($>$100:1 for Rs:Rp AND Tp:Ts) for best results.
\item[\textbf{P:}]  Needs precise remote control of orientations of any rotatable wave plates.
\item[\textbf{S:}]  Needs at least 2 single-photon detectors for the 2 measurement outputs or a time delay to time-multiplex the measurement outcomes onto a single detector.
\end{itemize}

Bob's mutually unbiased basis measurement system on the ground
\begin{itemize}
\item[\textbf{P:}]  Needs high-extinction ratio polarization beamsplitters ($>$100:1 for Rs:Rp AND Tp:Ts) for best results.
\item[\textbf{P:}]  Needs precise control of orientations of any rotatable wave plates.
\item[\textbf{S:}]  Needs at least 2 single-photon detectors for the 2 measurement outputs or a time delay to time-multiplex the measurement outcomes onto a single detector.
\item[\textbf{P:}]  Needs active polarization compensation to correct for rotations and phase shifts produced by the sending and receiving optics.
\item[\textbf{S:}]  Needs background-light filtering system.
\item[\textbf{S:}] Needs shared clock with satellite for coincidence post-processing or clever post-processing algorithm~\cite{wang2021synchronization}.
\end{itemize}
\subsection{Capability Requirements for HEQKD}
Given a system with a functioning polarization entanglement source and \emph{picosecond-pulsed mode-locked laser}, here are the unique system requirements to implement HEQKD in a quantum modem in space:

\begin{itemize}
\item[\textbf{T:}] \emph{Need fixed, unbalanced Mach-Zehnder interferometer (UMZI) for entanglement source pump (and a stabilization wavelength to send to the ground if the pump is not used for that)} \begin{itemize}
\item[\textbf{T:}] \emph{Needs to have delay be at least about 10}x \emph{longer than the pulse FWHM so there is no classical interference between the two paths of the UMZI.}
\item[\textbf{T:}] \emph{Needs to have delay be $> 3$}x\emph{ longer than detector FWHM timing jitter to have $< 1\%$ crosstalk rates (or misidentification errors).}
\item[\textbf{T:}] \emph{Needs to be polarization independent or use erasure polarizer at output.}
\item[\textbf{T:}] \emph{Needs to be phase stable, with respect to Alice’s measurement UMZI, within about ${5}^{\circ}$ or less for optimum performance.}
\item[\textbf{T:}] \emph{Will need pick-off of pump beam onto high-bandwidth photodiode or electrical synchronization signal from pump laser for reference clock to filter out the long-short + short-long paths from the long-long and short-short paths of a double UMZI interferometer(one for the pump, one for Alice/Bob).}
\end{itemize}
\item Alice's mutually unbiased basis measurement system in space
\begin{itemize}
\item[\textbf{T:}] \emph{Needs matched delay UMZI as pump UMZI.}
\item[\textbf{T:}] \emph{Needs to be phase stable, with respect to the pump UMZI, within about ${5}^{\circ}$ or less for optimum performance.}
\item[\textbf{P:}]  Needs high-extinction ratio polarization beamsplitters ($>$100:1 for Rs:Rp AND Tp:Ts) for best results.
\item[\textbf{T:}] \emph{Needs liquid crystals suitable for use in the environment (or make local environment suitable for them), as normally retardance is temperature sensitive.}
\item[\textbf{P/T:}] Needs precise remote control of \emph{voltages on all liquid crystals and} orientations of any rotatable wave plates.
\item[\textbf{S:}] Needs at least \emph{4} single-photon detectors for the \emph{4} measurement outputs.
\item[\textbf{T:}] \emph{Needs single-mode-fiber coupled outputs or multi-mode fiber coupled outputs and a robust multi-mode UMZI~\cite{PhysRevApplied.13.024047,PhysRevA.97.043847} to maintain time-bin entanglement quality.}
\end{itemize}
\item Bob's mutually unbiased basis measurement system on the ground
\begin{itemize}
\item[\textbf{T:}] \emph{Needs matched delay UMZI as pump UMZI with ability to change delay to compensate for Doppler shift.}
\item[\textbf{T:}] \emph{Needs to be phase stable, with respect to the pump UMZI, within about ${5}^{\circ}$ or less for optimum performance.}
\item[\textbf{T:}] \emph{Needs stabilization laser sent down from quantum modem in space to track Doppler shift.}
\item[\textbf{P:}]  Needs high-extinction ratio polarization beamsplitters ($>$100:1 for Rs:Rp AND Tp:Ts) for best results.
\item[\textbf{T:}] \emph{Needs liquid crystals suitable for use in the environment (or make local environment suitable for them), as normally retardance is temperature sensitive.}
\item[\textbf{P/T:}] Needs precise control of \emph{voltages on all liquid crystals and} orientations of any rotatable wave plates.
\item[\textbf{S:}] Needs at least \emph{4} single-photon detectors for the \emph{4} measurement outputs.
\item[\textbf{P:}]  Needs active polarization compensation to correct for rotations and phase shifts produced by the sending and receiving optics.
\item[\textbf{S:}] Needs background-light filtering system.
\item[\textbf{T:}] \emph{Needs adaptive-optics-aided single-mode fiber collection system or actively Doppler-compensating robust multi-mode interferometer to maintain time-bin entanglement quality.}
\item[\textbf{S:}] Needs shared clock with satellite for coincidence post-processing or clever post-processing algorithm~\cite{wang2021synchronization}.
\end{itemize}
\end{itemize}
\section{Time-bin Phase Stabilization and Calibration}\label{stab}
Due to natural environmental factors (vibration, temperature fluctuations, etc.), the phase between the time bins is prone to drift; to counteract this we stabilized the phase using an active proportional-integral (PI) feedback~\cite{PID} system. The phase was measured using some of the pump beam that was also sent (counter-propagating) through the analyzer interferometer in a mode that was vertically displaced, by about 15 mm, from the single-photon beam (see green lines in Fig. \ref{heqkdsetup}). The output of the stabilization interferometer was measured using Detectors $D_1$ and $D_2$, low-bandwidth amplified Si photodiodes (Thorlabs PDA36A) and a DAQ (NI USB-6210) to interface with the computer. Due to the stabilization beam wavelength not matching the design specification of some of the components, and because the low-bandwidth detectors could not distinguish the interfering time bins (e.g., the short (long) path in the pump interferometer and the long (short) path in Alice or Bob's analyzer interferometer) from the non-interfering ones (short paths in both interferometers and long paths in both interferometers), the visibility was quite low ($<10\%$) and also different for $D_1$ and $D_2$. This difference necessitated the use of a scaling factor $\gamma$ to equalize the amplitude of oscillation between $D_1$ and $D_2$. With this scaling factor, the error signal, $E$, used for the PI feedback algorithm was
\begin{equation}
E\equiv\frac{(I_{D_1}-\gamma{I_{D_2}})}{(I_{D_1}+\gamma{I_{D_2}})}\text{, with }\gamma\equiv0.6\text{.}
\end{equation}
The feedback system was designed to keep $E$ at zero by sending a signal to a piezo-actuated translation stage (Thorlabs 17.4-$\mu$m piezo AE0505D16F, Thorlabs TPZ001 piezo driver, and Newport 436 translation stage) holding the analyzer interferometer's right-angle prism to adjust the phase, at an update rate of 100 Hz. Independent PI systems were implemented for the Pump-Alice combined interferometer and for the Pump-Bob combined interferometer.

Additionally, for QKD it is not only necessary that the phase be kept stable, but also that it is calibrated to the correct value. By changing the relative phase between $\ket{H}$ and $\ket{V}$, the liquid crystals after Alice's interferometer were used to adjust the phase of states in basis 2 and 4 so that the photons were routed to the correct detector. For basis 3 and basis 4, it was necessary to tilt (about the vertical axes) a QWP before the source to adjust the phase $\phi$ of the polarization-entangled state $\ket{HH}+e^{i\phi}\ket{VV}$, so that $\ket{D}$ and $\ket{A}$ were routed to the correct detectors on each side.

\section{Time-bin Sorting Circuit Operation}\label{tbs}
\begin{figure}
\centerline{\includegraphics[height=1.7in,width=6in]{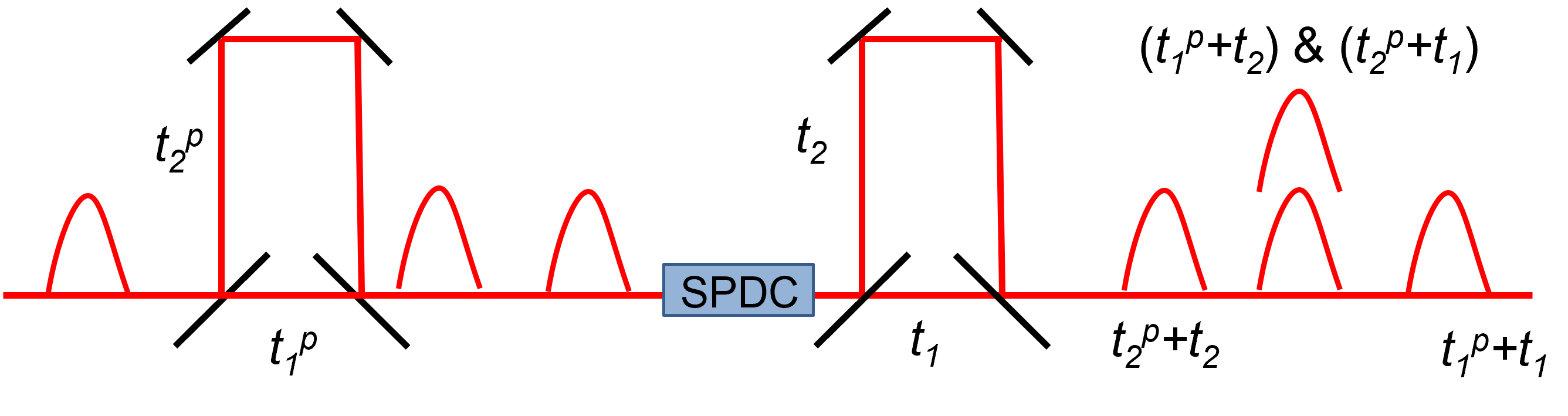}}\caption{Time-Bin Qubit Preparation and Measurement: This diagram illustrates how the time bins are created and what possible combinations of them exit the second delay-interferometer. Here we assume that $t_2-t_1=t_2^p-t_1^p$, i.e., that the path-length imbalances are matched. In this case, photons in either of the two middle time bins can interfere~\cite{TBQubit}.}\label{timebinpic}
\end{figure}

As displayed in Fig. \ref{timebinpic}, there are three time-bins which exit Alice or Bob's time-bin analyzer interferometer, each with a different exit time with respect to the pulse which entered the pump delay interferometer. It is imperative to be able to distinguish all three of these time bins for HEQKD; even for BBM92 we need to distinguish between two time-bins since those correspond to different polarization bases in our setup. The measurement of time-bin qubits, using free-running single-photon detectors alone, lacks the ability to sort basis measurements when events from different time bins are routed to the same detector, as in this experiment. We therefore developed a circuit which could filter the events corresponding to different time bins into different electrical signals. We used the pump laser as a clock reference and filtered the signals from each detector based on their delay with respect to the laser clock, using an AND gate with a window width of about $1$ ns. Each time bin has a unique delay with respect to the laser clock (CLK), so this enabled complete filtering of the time bins.

In the circuit, each detector output was copied (using ON Semi. NB7VQ14M) three times and was ANDed with a copy of the laser CLK that was delayed by the correct amount so that only events from one of the time bins was successfully transmitted through the AND gate. This process was executed for all eight detectors (four for Alice and four for Bob) and for all three time-bins, creating 24 unique output signals (12 for Alice and 12 for Bob). 

\begin{figure}
\centerline{\subfloat[]{{\includegraphics[scale=0.4]{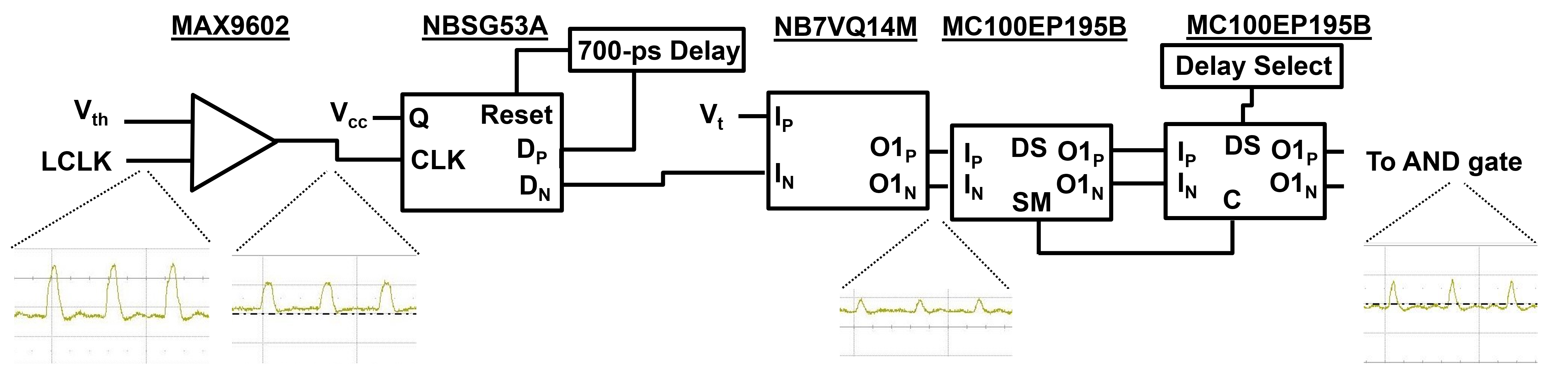}}}}\qquad\\
\centerline{\subfloat[]{{\includegraphics[scale=0.4]{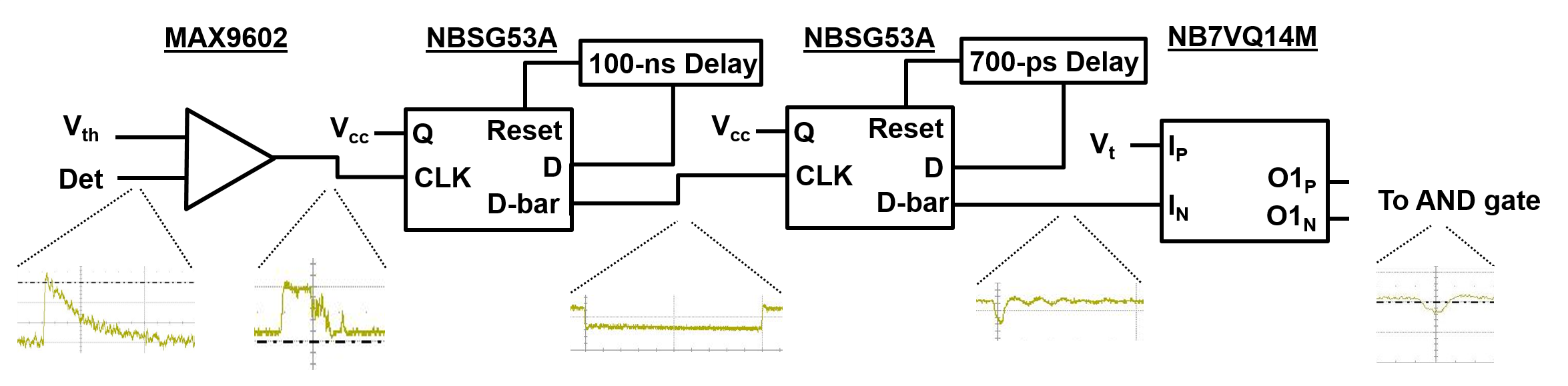}}}}
\caption{Descriptive Schematics: (a) laser CLK signal conditioning, using MAX 9602 high-speed comparator, a NBSG53A high-speed D-flip flops with about a $ 700$-ps delay, followed by an NB7VQ14M signal fanout for signal copying (only one out of four of the differential outputs is shown O1$_{\text{P,N}}$), and then two MC100EP195B precision delay chips. (b) Detector signal conditioning, using the MAX9602 high-speed comparator, several NBSG53A high-speed D-flip flops with about a $ 100$-ns delay, and then about a $ 700$-ps delay, followed by an NB7VQ14M signal fanout for signal copying (only one out of four of the differential outputs is shown O1$_{\text{P,N}}$). Signal traces shown for SNSPD input.  For an APD-detector signal, we used about a $ 45$-ns delay instead of about a $ 100$-ns delay for an SNSPD-detector signal, as shown above. V$_\text{th}$ = comparator voltage threshold; V$_\text{cc}$ = positive supply voltage; V$_\text{t}$ = The\'venin equivalent voltage; Det = signal from detector; LCLK = laser CLK. See product datasheets for definitions of other labels.}\label{fastAND}
\end{figure}

To use an AND gate (Analog Devices HMC746LC3C) and adjustable delay chip (ON Semi. MC100EP195B) with low jitter ($<100$ ps) and high bandwidth ($>1$ GHz), we needed a high-speed comparator (Maxim Int. MAX9602) to transform the electrical signals from the detectors (TTL and decaying exponential pulses) into signals compatible with high-speed differential logic standards like CML and PECL. Additionally, to achieve a sub-nanosecond pulse so the AND gate had about $1$-ns acceptance window, the pulses from the detectors were shortened to $<1$ ns using a cascade of two high-speed D flip-flops (ON Semi. NBSG53A). A pulse shortening effect was created by sending the comparator output into the CLK of the D flip-flop with Q (input) attached to logic HIGH and using a delayed copy of the output as a reset signal after the pulse was sent through the flip-flop. This long pulse was then sent through another D flip-flop with a short reset signal, producing a much shorter pulse (about $1$ ns) than the detector output pulses (about $40$ ns). See Fig. \ref{fastAND} for a pictorial description.

After the signal conditioning, the eight detector signals were copied three times (once for each time bin) and sent to the AND gate (Analog Devices HMC746LC3C). The laser CLK was copied 24 times and each copy was sent through two adjustable delay chips (ON Semi. MC100EP195B) in succession before being sent to the AND gate to be compared with the detector signal. At each AND gate, if the detector signal and laser CLK signal arrived within about 1 ns of one another, the output of the AND gate went HIGH for the duration of their overlap.

\section{Event Timetagging}
The detection events of Alice and Bob were recorded using separate time-tagging electronic devices (UQDevices UQD-Logic-16, with time-bin width of 156 ps), synchronized via a common 10-MHz sine-wave clock (Agilent 33250A Function Generator). Also, one channel on each time tagger was connected to a TTL pulse source (National Instruments DAQ USB-6210) through identical-length cables; this allowed the different time offsets of each time tagger to be measured accurately and subtracted out. The coincidence matrices for BBM92 and HEQKD are in Table \ref{BBM92detmapping} and Table \ref{HEQKDdetmapping}, respectively, indicating what states are measured in a given time bin, by a given detector pair.
%\begin{table}
%\caption{BBM92 Coincidence Matrix: Coincidence matrix of all detector and time-bin combinations used in the BBM92 protocol demonstration. See Fig. \ref{timebinpic} for pictorial definition of time bins $t_1^p+t_1$, ($t_1^p+t_2 \text{ \& }t_2^p+t_1$), and $t_2^p+t_2$.}
%\centerline{\includegraphics[angle=90,scale=0.2]{BBM92coincmatv5fppt.png}}
%\label{BBM92detmapping}
%\end{table}

\begin{sidewaystable}
\caption{BBM92 Coincidence Matrix: Coincidence matrix of all detector and time-bin combinations used in the BBM92 protocol demonstration. See Fig. \ref{timebinpic} for pictorial definition of time bins $t_1^p+t_1$, $(t_1^p+t_2)$ \& $(t_2^p+t_1)$, and $t_2^p+t_2$. The different colors represent the \color[rgb]{0,0.688,0.313}{H/V basis}\color{black}{,} \color[rgb]{0.75,0,0}{D/A basis}\color{black}{, and} \color[rgb]{0,0.438,0.75}{Different bases}\color{black}{, respectively}.}
\label{BBM92detmapping}

\centering
	\begin{tabular}{c| c| c c c c| c c c c}
	\hline
	\hline
 & & \multicolumn{4}{c}{$(t^p_1+t_1$)}&\multicolumn{4}{|c}{$(t^p_2+t_1)$ \& $(t^p_1+t_2)$}  \\ \hline
 & & $A_1$ & $A_2$ & $A_3$ & $A_4$ & $A_1$ & $A_2$ & $A_3$ & $A_4$ \\ \hline
\multirow{4}{*}{\rotatebox[origin=c]{-90}{$t^p_1+t_1$}}& $B_1$ &\color[rgb]{0,0.688,0.313}{$\ket{Ht^p_1t_1}\ket{Ht^p_1t_1}$ }&\color[rgb]{0,0.688,0.313}{$\ket{Ht^p_1t_1}\ket{Ht^p_1t_1}$ }&\color[rgb]{0,0.688,0.313}{$\ket{Ht^p_1t_1}\ket{Vt^p_1t_1}$ }&\color[rgb]{0,0.688,0.313}{$\ket{Ht^p_1t_1}\ket{Vt^p_1t_1}$ }&\color[rgb]{0,0.438,0.75}{$\ket{Ht^p_1t_1}\ket{At^p_1t_2}$}&\color[rgb]{0,0.438,0.75}{$\ket{Ht^p_1t_1}\ket{At^p_1t_2}$}&\color[rgb]{0,0.438,0.75}{$\ket{Ht^p_1t_1}\ket{Dt^p_1t_2}$}&\color[rgb]{0,0.438,0.75}{$\ket{Ht^p_1t_1}\ket{Dt^p_1t_2}$ }\\
&$B_2$& \color[rgb]{0,0.688,0.313}{$\ket{Ht^p_1t_1}\ket{Ht^p_1t_1}$ }&\color[rgb]{0,0.688,0.313}{ $\ket{Ht^p_1t_1}\ket{Ht^p_1t_1}$ }&\color[rgb]{0,0.688,0.313}{ $\ket{Ht^p_1t_1}\ket{Vt^p_1t_1}$ }&\color[rgb]{0,0.688,0.313}{ $\ket{Ht^p_1t_1}\ket{Vt^p_1t_1}$ }&\color[rgb]{0,0.438,0.75}{ $\ket{Ht^p_1t_1}\ket{At^p_1t_2}$ }&\color[rgb]{0,0.438,0.75}{ $\ket{Ht^p_1t_1}\ket{At^p_1t_2}$ }&\color[rgb]{0,0.438,0.75}{ $\ket{Ht^p_1t_1}\ket{Dt^p_1t_2}$ }&\color[rgb]{0,0.438,0.75}{ $\ket{Ht^p_1t_1}\ket{Dt^p_1t_2}$ }\\
&$B_3$& \color[rgb]{0,0.688,0.313}{$\ket{Vt^p_1t_1}\ket{Ht^p_1t_1}$ }&\color[rgb]{0,0.688,0.313}{ $\ket{Vt^p_1t_1}\ket{Ht^p_1t_1}$ }&\color[rgb]{0,0.688,0.313}{ $\ket{Vt^p_1t_1}\ket{Vt^p_1t_1}$ }&\color[rgb]{0,0.688,0.313}{ $\ket{Vt^p_1t_1}\ket{Vt^p_1t_1}$ }&\color[rgb]{0,0.438,0.75}{ $\ket{Vt^p_1t_1}\ket{At^p_1t_2}$ }&\color[rgb]{0,0.438,0.75}{ $\ket{Vt^p_1t_1}\ket{At^p_1t_2}$ }&\color[rgb]{0,0.438,0.75}{ $\ket{Vt^p_1t_1}\ket{Dt^p_1t_2}$ }&\color[rgb]{0,0.438,0.75}{ $\ket{Vt^p_1t_1}\ket{Dt^p_1t_2}$}\\
&$B_4$&\color[rgb]{0,0.688,0.313}{ $\ket{Vt^p_1t_1}\ket{Ht^p_1t_1}$ }&\color[rgb]{0,0.688,0.313}{ $\ket{Vt^p_1t_1}\ket{Ht^p_1t_1}$ }&\color[rgb]{0,0.688,0.313}{ $\ket{Vt^p_1t_1}\ket{Vt^p_1t_1}$ }&\color[rgb]{0,0.688,0.313}{ $\ket{Vt^p_1t_1}\ket{Vt^p_1t_1}$ }&\color[rgb]{0,0.438,0.75}{ $\ket{Vt^p_1t_1}\ket{At^p_1t_2}$ }&\color[rgb]{0,0.438,0.75}{ $\ket{Vt^p_1t_1}\ket{At^p_1t_2}$ }&\color[rgb]{0,0.438,0.75}{ $\ket{Vt^p_1t_1}\ket{Dt^p_1t_2}$ }&\color[rgb]{0,0.438,0.75}{ $\ket{Vt^p_1t_1}\ket{Dt^p_1t_2}$}\\\hline
\multirow{4}{*}{\rotatebox[origin=c]{-90}{\parbox[c]{1.5cm}{\centering $(t^p_2+t_1)$\& $(t^p_1+t_2)$}}} & $B_1$ &\color[rgb]{0,0.438,0.75}{ $\ket{At^p_1t_2}\ket{Ht^p_1t_1}$ }&\color[rgb]{0,0.438,0.75}{ $\ket{At^p_1t_2}\ket{Ht^p_1t_1}$ }&\color[rgb]{0,0.438,0.75}{ $\ket{At^p_1t_2}\ket{Vt^p_1t_1}$ }&\color[rgb]{0,0.438,0.75}{ $\ket{At^p_1t_2}\ket{Vt^p_1t_1}$ }&\color[rgb]{0.75,0,0}{ $\ket{At^p_1t_2}\ket{At^p_1t_2}$}&\color[rgb]{0.75,0,0}{ $\ket{At^p_1t_2}\ket{At^p_1t_2}$}&\color[rgb]{0.75,0,0}{ $\ket{At^p_1t_2}\ket{Dt^p_1t_2}$}&\color[rgb]{0.75,0,0}{ $\ket{At^p_1t_2}\ket{Dt^p_1t_2}$}\\
&$B_2$&\color[rgb]{0,0.438,0.75}{ $\ket{At^p_1t_2}\ket{Ht^p_1t_1}$ }&\color[rgb]{0,0.438,0.75}{ $\ket{At^p_1t_2}\ket{Ht^p_1t_1}$ }&\color[rgb]{0,0.438,0.75}{ $\ket{At^p_1t_2}\ket{Vt^p_1t_1}$ }&\color[rgb]{0,0.438,0.75}{ $\ket{At^p_1t_2}\ket{Vt^p_1t_1}$ }&\color[rgb]{0.75,0,0}{ $\ket{At^p_1t_2}\ket{At^p_1t_2}$}&\color[rgb]{0.75,0,0}{ $\ket{At^p_1t_2}\ket{At^p_1t_2}$}&\color[rgb]{0.75,0,0}{ $\ket{At^p_1t_2}\ket{Dt^p_1t_2}$}&\color[rgb]{0.75,0,0}{ $\ket{At^p_1t_2}\ket{Dt^p_1t_2}$}\\
&$B_3$&\color[rgb]{0,0.438,0.75}{ $\ket{Dt^p_1t_2}\ket{Ht^p_1t_1}$ }&\color[rgb]{0,0.438,0.75}{ $\ket{Dt^p_1t_2}\ket{Ht^p_1t_1}$ }&\color[rgb]{0,0.438,0.75}{ $\ket{Dt^p_1t_2}\ket{Vt^p_1t_1}$ }&\color[rgb]{0,0.438,0.75}{ $\ket{Dt^p_1t_2}\ket{Vt^p_1t_1}$ }&\color[rgb]{0.75,0,0}{ $\ket{Dt^p_1t_2}\ket{At^p_1t_2}$}&\color[rgb]{0.75,0,0}{ $\ket{Dt^p_1t_2}\ket{At^p_1t_2}$}&\color[rgb]{0.75,0,0}{ $\ket{Dt^p_1t_2}\ket{Dt^p_1t_2}$}&\color[rgb]{0.75,0,0}{ $\ket{Dt^p_1t_2}\ket{Dt^p_1t_2}$}\\
&$B_4$&\color[rgb]{0,0.438,0.75}{ $\ket{Dt^p_1t_2}\ket{Ht^p_1t_1}$ }&\color[rgb]{0,0.438,0.75}{ $\ket{Dt^p_1t_2}\ket{Ht^p_1t_1}$ }&\color[rgb]{0,0.438,0.75}{ $\ket{Dt^p_1t_2}\ket{Vt^p_1t_1}$ }&\color[rgb]{0,0.438,0.75}{ $\ket{Dt^p_1t_2}\ket{Vt^p_1t_1}$ }&\color[rgb]{0.75,0,0}{ $\ket{Dt^p_1t_2}\ket{At^p_1t_2}$}&\color[rgb]{0.75,0,0}{ $\ket{Dt^p_1t_2}\ket{At^p_1t_2}$}&\color[rgb]{0.75,0,0}{ $\ket{Dt^p_1t_2}\ket{Dt^p_1t_2}$}&\color[rgb]{0.75,0,0}{ $\ket{Dt^p_1t_2}\ket{Dt^p_1t_2}$}\\
	\hline
	\hline		
			\end{tabular}

\end{sidewaystable}

\begin{sidewaystable}
\caption{HEQKD Coincidence Matrix: Coincidence matrix of all detector and time bin combinations used in the HEQKD protocol demonstration. See Fig. \ref{timebinpic} for pictorial definition of time bins $t_1^p+t_1$, $(t_1^p+t_2)$ \& $(t_2^p+t_1)$, and $t_2^p+t_2$. The different colors represent the \color[rgb]{0,0.688,0.313}{Extremal basis}\color{black}{,} \color[rgb]{0.75,0,0}{Middle basis}\color{black}{, and} \color[rgb]{0,0.438,0.75}{Different bases}\color{black}{, respectively}.}
\begin{adjustbox}{scale=0.34}
\centering
\begin{tabular}{c|c| c| c c c c| c c c c| c c c c}
	\hline
	\hline
&&&	\multicolumn{12}{c}{\centering Basis 1 or 2}\\ \hline
 &&& \multicolumn{4}{c}{$(t^p_1+t_1$)}&\multicolumn{4}{|c|}{$(t^p_2+t_1)$ \& $(t^p_1+t_2)$}& \multicolumn{4}{c}{$(t^p_2+t_2$)}  \\ \hline
%\rotatebox[origin=c]{-90}{\parbox[c]{1cm}{\centering Basis 1 or 2}} 
&&& $A_1$ & $A_2$ & $A_3$ & $A_4$ & $A_1$ & $A_2$ & $A_3$ & $A_4$& $A_1$ & $A_2$ & $A_3$ & $A_4$ \\ \hline
\multirow{12}{*}{\rotatebox[origin=c]{-90}{\centering Basis 1 or 2}}&\multirow{4}{*}{\rotatebox[origin=c]{-90}{$t^p_1+t_1$}}& $B_1$ &\color[rgb]{0,0.688,0.313}{$\ket{0}\ket{0}$ }&\color[rgb]{0,0.688,0.313}{$\ket{0}\ket{0}$ }&\color[rgb]{0,0.688,0.313}{$\ket{0}\ket{2}$ }&\color[rgb]{0,0.688,0.313}{$\ket{0}\ket{2}$ }&\color[rgb]{0,0.438,0.75}{$\ket{0}\frac{1}{\sqrt{2}}(\ket{2}+\ket{3})$ }&\color[rgb]{0,0.438,0.75}{$\ket{0}\frac{1}{\sqrt{2}}(\ket{2}-\ket{3})$ }&\color[rgb]{0,0.438,0.75}{$\ket{0}\frac{1}{\sqrt{2}}(\ket{0}+\ket{1})$ }&\color[rgb]{0,0.438,0.75}{$\ket{0}\frac{1}{\sqrt{2}}(\ket{0}-\ket{1})$
}&\color[rgb]{0,0.688,0.313}{$\ket{0}\ket{1}$
}&\color[rgb]{0,0.688,0.313}{$\ket{0}\ket{1}$
}&\color[rgb]{0,0.688,0.313}{$\ket{0}\ket{3}$
}&\color[rgb]{0,0.688,0.313}{$\ket{0}\ket{3}$}\\
&&$B_2$&\color[rgb]{0,0.688,0.313}{$\ket{0}\ket{0}$ }&\color[rgb]{0,0.688,0.313}{$\ket{0}\ket{0}$ }&\color[rgb]{0,0.688,0.313}{$\ket{0}\ket{2}$ }&\color[rgb]{0,0.688,0.313}{$\ket{0}\ket{2}$ }&\color[rgb]{0,0.438,0.75}{$\ket{0}\frac{1}{\sqrt{2}}(\ket{2}+\ket{3})$ }&\color[rgb]{0,0.438,0.75}{$\ket{0}\frac{1}{\sqrt{2}}(\ket{2}-\ket{3})$ }&\color[rgb]{0,0.438,0.75}{$\ket{0}\frac{1}{\sqrt{2}}(\ket{0}+\ket{1})$ }&\color[rgb]{0,0.438,0.75}{$\ket{0}\frac{1}{\sqrt{2}}(\ket{0}-\ket{1})$
}&\color[rgb]{0,0.688,0.313}{$\ket{0}\ket{1}$
}&\color[rgb]{0,0.688,0.313}{$\ket{0}\ket{1}$
}&\color[rgb]{0,0.688,0.313}{$\ket{0}\ket{3}$
}&\color[rgb]{0,0.688,0.313}{$\ket{0}\ket{3}$}\\
\hline
\hline
			\end{tabular}
			
\end{adjustbox}

%\centerline{\includegraphics[angle=90,height=7.5in,width=1.18in]{Bases1212coincmatv7.png}\includegraphics[angle=90,height=7.5in,width=1.18in]{Bases1234coincmatv7.png}\includegraphics[angle=90,height=7.5in,width=1.18in]{Bases3412coincmatv7.png}\includegraphics[angle=90,height=7.5in,width=1.18in]{Bases3434coincmatv8.png}}
\label{HEQKDdetmapping}

\end{sidewaystable}

\clearpage
%\bibliography{QKD2021}{}

\end{document}